\documentclass[11pt]{article}
\usepackage{amsmath,amssymb,color,epsf}

\makeatletter
\@addtoreset{equation}{section}
\makeatother

\newcommand{\hoch}[1]{$\, ^{#1}$}

\newcommand{\be}{\begin{equation}}
\newcommand{\ee}{\end{equation}}
\newcommand{\bea}{\setlength\arraycolsep{2pt} \begin{eqnarray}}
\newcommand{\eea}{\end{eqnarray}}
\newcommand{\nn}{\nonumber}

\def\ft#1#2{{\textstyle{\frac{\scriptstyle #1}{\scriptstyle #2} } }}
\def\fft#1#2{{\frac{#1}{#2}}}

\def\0{{\sst{(0)}}}
\def\1{{\sst{(1)}}}
\def\2{{\sst{(2)}}}
\def\3{{\sst{(3)}}}
\def\4{{\sst{(4)}}}
\def\5{{\sst{(5)}}}
\def\6{{\sst{(6)}}}
\def\7{{\sst{(7)}}}
\def\8{{\sst{(8)}}}
\def\sst#1{{\scriptscriptstyle #1}}

\def\cG{{{\cal G}}}

\begin{document}

\begin{flushright}
\hfill{ \
MIFPA-12-10\ \ \ \ }
\end{flushright}

\begin{center}
{\large {\bf AdS and Lifshitz Black Holes in Conformal and Einstein-Weyl Gravities}}

\vspace{10pt}

H. L\"u\hoch{1,2}, Yi Pang\hoch{3}, C.N. Pope\hoch{3,4} and J.F.
V\'azquez-Poritz\hoch{5,6}

\vspace{10pt}

\hoch{1}{\it China Economics and Management Academy\\
Central University of Finance and Economics, Beijing 100081}

\vspace{10pt}

\hoch{2}{\it Institute for Advanced Study, Shenzhen
University, Nanhai Ave 3688, Shenzhen 518060}

\vspace{10pt}

\hoch{3} {\it George P. \& Cynthia Woods Mitchell  Institute
for Fundamental Physics and Astronomy,\\
Texas A\&M University, College Station, TX 77843, USA}

\vspace{10pt}

\hoch{4}{\it DAMTP, Centre for Mathematical Sciences,
 Cambridge University,\\  Wilberforce Road, Cambridge CB3 OWA, UK}

\vspace{10pt}

\hoch{5}{\it New York City College of Technology, The City
University of New York\\ 300 Jay Street, Brooklyn NY 11201, USA}

\vspace{10pt}

\hoch{6}{\it The Graduate School and University Center, The City
University of New York\\ 365 Fifth Avenue, New York NY 10016, USA}

\vspace{20pt}

\underline{ABSTRACT}
\end{center}

We study black hole solutions in extended gravities with
higher-order curvature terms, including conformal and Einstein-Weyl
gravities.  In addition to the usual AdS vacuum, the theories admit
Lifshitz and Schr\"odinger vacua. The AdS black hole in conformal
gravity contains an additional parameter over and above the mass,
which may be interpreted as a massive spin-2 hair.  By considering
the first law of thermodynamics, we find that it is necessary to
introduce an associated additional intensive/extensive pair of
thermodynamic quantities. We also obtain new Liftshitz black holes
in conformal gravity and study their thermodynamics.  We use a
numerical approach to demonstrate that AdS black holes beyond the
Schwarzschild-AdS solution exist in Einstein-Weyl gravity. We also
demonstrate the existence of asymptotically Lifshitz black holes in
Einstein-Weyl gravity.  The Lifshitz black holes arise at the
boundary of the parameter ranges for the AdS black holes. Outside
the range, the solutions develop naked singularities. The
asymptotically AdS and Lifshitz black holes provide an interesting
phase transition, in the corresponding boundary field theory, from a
relativistic Lorentzian system to a non-relativistic Lifshitz
system.

\vspace{15pt}

\thispagestyle{empty}

\pagebreak
\voffset=-40pt
\setcounter{page}{1}

\tableofcontents

\addtocontents{toc}{\protect\setcounter{tocdepth}{2}}


\newpage

\section{Introduction}

   Theories of gravity extended by the addition of higher-order
curvature terms are of interest for a number of reasons. One
motivation is to investigate whether suitably extended
four-dimensional gravity can be quantized in its own right. It has
been shown that Einstein gravity extended by the inclusion of
quadratic curvature terms is perturbatively renormalizable
\cite{stelle1}.  However, the price to be paid for achieving
renormalizability is that the theory then contains massive spin-2
modes as well as the massless graviton and, furthermore, that the
massive modes are ghostlike (i.e. their kinetic term has the wrong
sign). Three-dimensional models of gravity, for which the UV
divergence problems are inherently less severe, have also been
studied extensively. While Einstein gravity itself is essentially
trivial in three dimensions, extensions to include higher-order
derivative terms lead to interesting toy models with dynamical
content and the possibility of well-controlled UV behaviour.  Such
extensions in three dimensions include topologically massive gravity
\cite{tmg1}, and more recently, ``New Massive Gravity'' \cite{nmg}.
The theory can be rendered ghost free, and equivalent to a theory
with a standard Einstein-Hilbert action, after truncating out modes
with logarithmic fall-off by imposing an appropriate boundary
condition of AdS$_3$.  (See, for example, \cite{lss}.)
Supersymmetric extensions were considered in
\cite{abrhst}-\cite{lphybrid}.

The situation is rather more subtle in four dimensions. An analogous
``critical gravity'' in four dimensions was considered in
\cite{lpcritical}. The Lagrangian consists of the Einstein-Hilbert
term with a cosmological constant $\Lambda$ and an additional
higher-order term proportional to the square of the Weyl tensor,
with a coupling constant $\alpha$. It was shown that there is a
critical relation between $\alpha$ and $\Lambda$ for which the
generically-present massive spin-2 modes disappear, and are instead
replaced by modes with a logarithmic fall-off \cite{d4log} (see also
\cite{aflog,ggst}).  These log modes are ghostlike in nature
\cite{prghost,Liu:2011kf}, but since they fall off more slowly than
do the massless spin-2 modes, they can be truncated out by imposing
an appropriate AdS boundary condition. The resulting theory is then
somewhat trivial, in the sense that the remaining massless graviton
has zero on-shell energy. Furthermore, the mass and the entropy of
standard Schwarzschild-AdS black holes are both zero in the critical
theory. Analogous critical theories exist also in higher-dimensional
extended gravities with curvature-squared terms \cite{dllpst}.

In fact, it was observed in \cite{Lu:2011ks} that one can generalise
critical gravity to a wider class of Weyl-squared extensions to
cosmological Einstein gravity, where $\alpha$ does not take the
critical value. For a certain range of values for $\alpha$, the
mass-squared of the massive spin-2 mode in the AdS$_4$ background is
negative, but not sufficiently negative to imply tachyonic
behaviour.  However, this mode, which is again ghostlike, has a
slower fall-off than the massless graviton and so it can be
truncated by imposing appropriate AdS boundary conditions. This
extended class of models has been investigated further in
\cite{Bergshoeff:2011xy}-\cite{Nutma:2012ss}.

One possible approach would be to begin with the
conformally-invariant theory described by a pure Weyl-squared
action. Being the local gauge theory of the conformal group, this
theory of ``conformal gravity'' has the virtue of yielding a
convergent Euclidean functional integral, and also of
renormalizability and asymptotic freedom \cite{Fradkin:1981iu}. One
might then argue \cite{Adler:1982ri, Zee:1980sj} that quantum
fluctuations would break the scale invariance, and thereby generate
an Einstein-Hilbert term in the low-energy effective action. Thus,
the Einstein-Weyl extensions of critical gravity described above
effectively describe the emergence \cite{mald} of Einstein gravity
from conformal gravity. It then becomes of interest to investigate
the classical solutions of conformal gravity and Einstein-Weyl
gravity. Any solution of Einstein gravity with a cosmological
constant is also a solution of Einstein-Weyl gravity. However, the
Weyl-squared term gives rise to fourth-order equations of motion,
which are highly non-linear, and so it is in general rather
difficult to find the further new solutions that exist over and
above the pure Einstein solutions. One of the main purposes of the
present paper is to search for such new solutions under the
simplifying assumption of spherical symmetry (and certain
generalisations of this).

The investigation of solutions in higher-derivative extensions of
Einstein gravity is also of interest from the AdS/CFT viewpoint, not
least because it is known that such higher-order terms arise in
string theory. Furthermore, although originally the AdS/CFT
correspondence was conceived as a duality between a conformal field
theory and a string theory, the idea of holography has been
generalized to broader classes of gauge/gravity duality outside the
string theoretical context.

Recently, holographic techniques have been used to study
non-relativistic systems, such as atomic gases at ultra-low
temperature. This entails two types of gravitational backgrounds:
those which correspond to Lifshitz-like fixed points
\cite{Kachru:2008yh} and Schr\"odinger-like fixed points
\cite{Son:2008ye, Balasubramanian:2008dm}. In the context of
condensed matter theory, various systems exhibit a dynamical scaling
near fixed points:
\begin{equation}\label{scaling}
t\rightarrow \lambda^z t,\qquad x_i\rightarrow \lambda x_i,\qquad
z\neq1\,.
\end{equation}
In other words, rather than obeying the conformal scale invariance
$t\rightarrow \lambda t$, $x_i\rightarrow \lambda x_i$, the temporal and
the spatial coordinates scale anisotropically.

Requiring also time and space translation invariance, spatial
rotational symmetry, spatial parity and time reversal invariance,
the authors of \cite{Kachru:2008yh} were led to consider $D$
dimensional geometries of the form
\begin{equation}\label{L-metric}
    ds^2=\ell^2\biggl(-r^{2z}dt^2+r^2dx_idx^i+\frac{dr^2}{r^2}\biggr)\,.
\end{equation}
This metric obeys the scaling relation (\ref{scaling}) if one also
scales $r\rightarrow\lambda^{-1}r$. If $z=1$, the metric reduces to
the usual AdS metric in Poincar\'e coordinates with AdS radius
$\ell$. Metrics of the form (\ref{L-metric}) can be obtained as
solutions in general relativity with a negative cosmological
constant if appropriate matter is included. For example, solutions
were found by introducing 1-form and 2-form gauge fields
\cite{Kachru:2008yh}; a massive vector field \cite{Taylor:2008tg};
in an abelian Higgs model \cite{Gubser:2009cg}; and with a charged
perfect fluid \cite{Hartnoll:2010gu}. A class of Lifshitz black hole
solutions with non-planar horizons was found in
\cite{Danielsson:2009gi,Mann:2009yx}.  String theory and
supergravity embeddings have been found in
\cite{Hartnoll}-\cite{singh2}.

In a similar vein, $D$-dimensional geometries which exhibit
Schr\"odinger symmetry are described by a metric of the form
\cite{Son:2008ye, Balasubramanian:2008dm}
\begin{equation}\label{S-metric}
ds^2=\ell^2\biggl(-r^{2z}dt^2+r^2(-dt
d\xi+dx_idx^i)+\frac{dr^2}{r^2}\biggr)\,,
\end{equation}
which is conformally related to a pp-wave spacetime. This metric
obeys the scaling relation
\be
t\rightarrow \lambda^z t,\qquad x_i\rightarrow \lambda x_i,\qquad
r\rightarrow \lambda r,\qquad \xi\rightarrow \lambda^{2-z} \xi,
\qquad z\neq1\,. \ee
If momentum along the $\xi$ direction is interpreted as rest mass,
then this metric describes a system which exhibits time and space
translation invariance, spatial rotational symmetry, and invariance
under the combined operations of time reversal and charge
conjugation. These geometries have been embedded in string theory
\cite{Herzog:2008wg, Maldacena:2008wh}.

The organization of this paper is as follows.  Section 2 contains a
brief description of four-dimensional Einstein-Weyl gravity,
including the equations of motion. In section 3, we summarise some
salient features of the AdS$_4$ solution of Einstein-Weyl gravity
and the nature of the linearised fluctuations around the AdS$_4$
background.  We also discuss the Lifshitz solutions of Einstein-Weyl
gravity, deriving the relation between the coupling strength
$\alpha$ of the Weyl-squared term and the value of the Lifshitz
anisotropy parameter $z$.  We also find Schr\"odinger-type
solutions. In section 4, we consider a black hole type ansatz for
spherically-symmetric solutions of Einstein-Weyl gravity, and
generalisations where the spatial sections are flat or hyperbolic
instead of spherical. We show that the fourth-order equations of
motion can be reduced to second-order equations for the metric
functions.  In the special case of flat spatial sections, we also
derive a conserved Noether charge, which for the standard black hole
solution is related to the mass.

In section 5, we consider spherically-symmetric black holes in the
specific case of pure conformal gravity.  They can have either AdS
or Lifshitz asymptotic behaviour.  The AdS black holes in conformal
gravity have an additional parameter over and above the mass, and
this leads to interesting consequences when one considers their
thermodynamics.  We discuss how one may generalise the first law of
thermodynamics to include the additional parameter.  We also
consider the asymptotically-Lifshitz black holes in conformal
gravity, which can have either $z=4$ or $z=0$. In section 6, we
extend this discussion to Einstein-Weyl gravities.  Now, it appears
that the equations governing the metric functions for the
spherically-symmetric ansatz are too complicated to be solvable in
general, and so we have to resort to numerical methods in order to
go beyond the known Schwarzschild-AdS metrics.  To do this, we first
give a discussion of the forms of the solutions in the near-horizon
and the asymptotic regions. Then upon performing numerical
integrations outwards from the near-horizon region, we find
indications that more general black hole solutions do indeed exist,
at least when the $\alpha$ coupling parameter for the Weyl-squared
term lies in an appropriate range.  Section 7 contains our
conclusions. We present further solutions of conformal gravity and
general extended gravities with quadratic curvature-squared terms in
appendices A and B, respectively. We summarise some results on the
calculation of the mass for black holes in the critical theory and in
conformal gravity in Appendices C and D, respectively.

\section{Extended and Critical Gravity}

We begin by considering the action
\begin{equation}
I = \fft{1}{2\kappa^2}\,
  \int \sqrt{-g}\,d^4x(R -2\Lambda + \alpha R^{\mu\nu} R_{\mu\nu} +
   \beta R^2 + \gamma E_{\rm GB})\,,\label{action}
\end{equation}
where $\kappa^2 = 8\pi G$ and $E_{GB}$ is the Gauss-Bonnet term
\begin{equation}
E_{GB}=R^2 - 4 R^{\mu\nu}R_{\mu\nu} +
      R^{\mu\nu\rho\sigma}R_{\mu\nu\rho\sigma}\,.
\end{equation}
Although this term does not contribute to the equations of motion in
four dimensions, it can have non-trivial consequences for
thermodynamics in the higher-derivative theory, and so we shall
include it in our discussion.

The equations of motion that follow from the action (\ref{action})
are
\begin{equation}
  \cG_{\mu\nu} + E_{\mu\nu}=0\,,\label{eom}
\end{equation}
where
\begin{eqnarray}
\cG_{\mu\nu} &=& R_{\mu\nu} -\ft12 R_{\mu\nu} + \Lambda\, g_{\mu\nu}\,,
  \label{cGdef}\\
E_{\mu\nu} &=& 2\alpha(R_{\mu\rho}\, R_\nu{}^\rho -\ft14 R^{\rho\sigma}
  R_{\rho\sigma}\, g_{\mu\nu}) + 2\beta R\, (R_{\mu\nu} -\ft14 R\, g_{\mu\nu})
\nn\\
&&+\alpha\, ( \square R_{\mu\nu} + \nabla_\rho\nabla_\sigma R^{\rho\sigma}\,
  g_{\mu\nu} - 2\nabla_\rho \nabla_{(\mu} R_{\nu)}{}^\rho) +
 2\beta\, (g_{\mu\nu}\, \square R -\nabla_\mu\nabla_\nu R)\,.\label{Edef}
\end{eqnarray}

When $\beta=-\ft13\alpha$ and $\gamma=\ft12\alpha$, the theory
describes what we shall call Einstein-Weyl gravity, with the
action
\begin{equation}
I = \fft{1}{2\kappa^2}\,
  \int \sqrt{-g}\,d^4x(R -2\Lambda + \ft12\alpha |{\rm Weyl}|^2)\,,
\label{actionEW}
\end{equation}
where
\begin{equation}
|{\rm Weyl}|^2 = R^{\mu\nu\rho\sigma} R_{\mu\nu\rho\sigma} - 2
R^{\mu\nu}R_{\mu\nu} + \ft13 R^2\,.
\end{equation}
Note that the equations of motion following from this action can be
written as\footnote{In deriving this result it is helpful to note
that, in four dimensions, the Weyl tensor satisfies the identity
$C_{\mu\rho\sigma\lambda} C_\nu{}^{\rho\sigma\lambda} = \ft14
C_{\rho\sigma\lambda\tau} C^{\rho\sigma\lambda\tau}\, g_{\mu\nu}$.
This can be seen easily by employing 2-component spinor notation.}
\be
R_{\mu\nu} -\ft12 R  g_{\mu\nu} + \Lambda g_{\mu\nu}
-2\alpha(2\nabla^\rho \nabla^\sigma + R^{\rho\sigma})
C_{\mu\rho\sigma\nu}=0 \,.\label{einsteinweyleom} \ee

We shall also sometimes consider the limit of {\it conformal
gravity}, where only the Weyl-squared term in the action is
retained.  This can be described as the $\alpha\longrightarrow
\infty$ limit of the Einstein-Weyl action (\ref{actionEW}).  The
action of conformal gravity is conformally invariant, which implies
that the equations of motion determine the metric only up to an
arbitrary conformal factor.

A special feature of four-dimensional Einstein gravity with
curvature-squared terms is that any solution of the pure Einstein
theory continues to be a solution of the theory with the quadratic
modifications. Thus, in particular, the Schwarzschild-AdS black hole
solution
\be
ds^2 = -\Big(k - \fft{2m}{r} - \ft13\Lambda r^2\Big)\, dt^2 +
 \Big(k - \fft{2m}{r} - \ft13\Lambda r^2\Big)^{-1}\, dr^2 +
 r^2 d\Omega_{2,k}^2 \label{schwads}
\ee
of Einstein gravity is also a solution in the higher-derivative theory.
Here $d\Omega_{2,k}^2$ denotes the metric on a unit 2-sphere ($k=1$),
unit hyperbolic plane ($k=-1$) or 2-torus ($k=0$), and may be written as
\begin{equation}
d\Omega_{2,k}^2 =\fft{dx^2}{1-k\,x^2} + (1-k\,x^2) dy^2\,.
\end{equation}

\section{AdS, Lifshitz and Schr\"odinger Vacua}

Unlike the $D=3$ or $D\ge 5$ cases, for $D=4$ the cosmological
constant of the (A)dS vacuum is not modified by the quadratic
curvature terms, and hence we have only one such vacuum with
cosmological constant $\Lambda$. In this paper we shall consider
only negative $\Lambda$, and furthermore, without loss of
generality, we shall from now on set $\Lambda=-3$.

The linearised fluctuations around the AdS$_4$ vacuum in
Einstein-Weyl gravity were analyzed in \cite{lpcritical}.  It turns
out that the scalar trace mode decouples from the spectrum, which
then contains just massless and massive spin-2 modes, satisfying
\begin{equation}
\alpha (\Box +2)(\Box +2 -m^2)h_{\mu\nu}=0\,,
\end{equation}
where $h_{\mu\nu}$ is transverse traceless and
\begin{equation}
m^2 =-2 -\fft{1}{\alpha}\,.
\end{equation}

The characteristics of the linearised theory depend upon the value
of the parameter $\alpha$, and are summarised in the table below.

\bigskip\bigskip
\centerline{
\begin{tabular}{|c|c|c|c|c|c|c|c|}\hline
&$-\infty < \alpha <-\ft12$  & $\alpha=-\ft12$ & $-\ft12 <\alpha <0$
& $\alpha=0$ & $0 <\alpha <4$ & $4 \le\alpha<\infty $\\ \hline\hline
 & $-\ft94 \le m^2<0$ & $m^2=0$ & $m^2>0$ & -- & $m^2<-\ft94$
 & $-\ft94\le m^2 <0$ \\
{\bf Massive:} & & Ghost & Ghost & --& Ghost& Ghost\\
& &Log & & --&Tachyon & \\
& Trunc.& Trunc.& Non-trunc.& --& & Trunc. \\ \hline
{\bf Massless:}& Ghost& Null &&&&\\ \hline
\end{tabular}}
\bigskip

\noindent{Table 1: The characteristics of the massive and massless
spin-2 modes in Einstein-Weyl gravity for finite values of the
parameter $\alpha$. When not indicated to the contrary, the modes
are non-ghostlike.}
\bigskip

\noindent Owing to the fact that the background is AdS rather than
Minkowski spacetime, there is an allowed range of negative
mass-squared values for the massive mode, $-\ft94\le m^2<0$, for
which it is still non-tachyonic.  In this range, the massive mode
actually falls off less rapidly at infinity than the massless mode,
and so it can be truncated from the theory by imposing a suitable
boundary condition at infinity.  In the critical theory, which
occurs when $\alpha=-\ft12$, the massive mode becomes massless and
in fact a new type of mode with logarithmic fall off arises. The
usual massless mode has zero norm in the critical theory, and the
logarithmic mode is ghostlike. The logarithmic mode could be
truncated by imposing appropriate boundary conditions, but this
would leave only the zero-norm massless graviton
\cite{prghost,Liu:2011kf}. The case $\alpha=0$ corresponds to
ordinary Einstein gravity, in which case there is of course no
massive mode.  Not depicted in the table is the case
$\alpha=\pm\infty$, which corresponds to the pure Weyl-squared
conformal theory. In the conformal theory the massive mode has
$m^2=-2$, and so although negative, it is not tachyonic.

In addition to the AdS vacuum, the theory (\ref{action}) also admits
Lifshitz solutions, for which the metric is given by
\begin{equation}
ds^2=\fft{dr^2}{\sigma\,r^2} - r^{2z} dt^2 + r^2 (dx^2 + dy^2)\,,
\end{equation}
where
\begin{equation}
(z^2+2)\alpha + 2(z^2 + 2z + 3)\beta =\ft1{12} (z^2 + 2z+3)\,,\qquad
\sigma = \fft{6}{z^2+2z + 3}\,.\label{sigma}
\end{equation}
For the special case of Einstein-Weyl gravity, where $\beta=-\alpha/3$,
we have
\begin{equation}
\alpha=\fft{z^2+2z+3}{4z(z-4)}\,.\label{zeq}
\end{equation}
For conformal gravity, corresponding to $\alpha=\infty$, equation
(\ref{zeq}) implies that the Lifshitz scaling parameter $z$ can take the
values $z=4$ or $z=0$.
At the critical point, on the other hand, where $\alpha=-\fft12$,
both roots of (\ref{zeq}) give $z=1$.  For general values of $\alpha$
we have
\begin{equation}
z=\fft{8\alpha+1\pm\sqrt{2(1+2\alpha)(16\alpha-1)}}{4\alpha-1}\,.
\label{zsol}
\end{equation}
Thus, the reality of $z$ requires that $\alpha\ge \ft1{16}$ or $\alpha
\le \ft12$.

For conformal gravity, we find that there are also Lifshitz-like solutions
with $S^2$ or $H^2$ spatial topologies as well as $T^2$.  The metrics for
all three cases can be written as
\begin{equation}
ds^2 = -r^{2z} (1+\fft{k}{r^2}) dt^2+ \fft{4
dr^2}{r^2(1+\fft{k}{r^2})} + r^2
d\Omega_{2,k}^2\,,\label{conflifsvac}
\end{equation}
with $z=0$ and $4$.

Finally, we consider Schr\"odinger vacua, whose metric takes the
form
\begin{equation}
ds^2=-r^{2z} dt^2 + \fft{dr^2}{r^2} + r^2 (-2dt dx + dy^2)\,.
\end{equation}
For $z=1$ and $z=-\ft12$, the metrics are Einstein with
$\Lambda=-3$, and hence they are solutions for all $\alpha$ and
$\beta$.  In particular, the $z=1$ case is simply the AdS metric,
whilst if $z=-\ft12$ it is the Kaigorodov metric describing a
pp-wave propagating in AdS \cite{kaig,clpkaig}. In general, $z$
satisfies \cite{aflog}
\begin{equation}
1 - 24 \beta + \alpha (4z^2 - 2z - 8)=0\,.
\end{equation}
In Einstein-Weyl gravity, we have
\begin{equation}
\alpha=\fft{1}{2z(1-2z)}\,.
\end{equation}
In conformal gravity, $z$ can take values 1, $\ft12$, 0 or $-\ft12$.
Some asymptotic Schr\"odinger solutions are presented in appendix
A.1.

\section{Black Hole Ansatz and Equations of Motion}

\subsection{General equations for $k=1$, 0 or $-1$}

In this paper, we focus on the construction of static,
spherically-symmetric (or $H^2$ or $T^2$ symmetric) black hole
solutions that are asymptotic to either the AdS or the Lifshitz
vacua discussed in the previous section. We may therefore, without
loss of generality, consider the ansatz
\begin{equation}
ds^2=-a(r)\, dt^2 + \fft{dr^2}{f(r)} + r^2
d\Omega_{2,k}^2\,.\label{bhansatz1}
\end{equation}
The equations of motion for $a$ and $f$ may be derived from the
Lagrangian obtained by substituting this ansatz into the action
(\ref{action}).  Since we are interested specifically in the case of
Einstein-Weyl gravity, where $\beta=-\ft13\alpha$, and since the
equations of motion are greatly simplified in this
case,\footnote{The reason for the simplification is that the trace
of the Weyl-squared contribution to the equations of motion vanishes
(see equation (\ref{einsteinweyleom})), and so this projection is
identical in Einstein gravity and Einstein-Weyl gravity.} we shall
present the results under this specialisation.  We then find that
the equations can be reduced to the second-order system
\begin{equation}
a''=\fft{r^2f a'^2 + 4a^2(k+6r^2-f-rf') - ra a'(4f+rf')}{2r^2af}\,,
\label{trace}
\end{equation}
and
\begin{eqnarray}
f'' &=& \fft{1}{2 r^2 a^2 f (r a'-2a)}\Big(\fft{4r^2a^2}{\alpha}
(a(k+3r^2-f) - r f a') +r^3 f^2 a'^3 + 2r^2 a^2 fa'(8r-f')\cr
&&-r^2afa'^2(3f+r f') - a^3 (48r^4-16r^2f + 8f^2-24r^3f' + 4rff'+
3r^2f'^2)\cr &&-4ka^3(4r^2-2f-r f')\Big)\,.\label{f2peom}
\end{eqnarray}

\subsection{Conserved Noether charge for $k=0$ case}

In the case of a toroidal spatial geometry (i.e. $k=0$), the system
of equations has an additional global symmetry, and hence there is
an associated conserved Noether charge.  In order to discuss this,
it is helpful temporarily to choose a different parameterisation of
the metric ansatz (\ref{bhansatz1}), using a new radial coordinate
$\rho$ such that $r^2=b(\rho)$ and $dr^2/f=ab^2h d\rho^2$, so that
the metric is now written as
\be
ds^2 = - a(\rho) dt^2 + a(\rho) b(\rho)^2\, h(\rho) d\rho^2 + b(\rho)\,
  dx^i dx^i\,.
\ee
Since the additional global symmetry arises regardless of whether
or not we choose the Weyl-squared combination $\beta=-\ft13\alpha$,
we shall keep these two parameters arbitrary in the following
discussion. Substituting into the action (\ref{action}) yields an
effective Lagrangian for $a$, $b$ and $h$.  The function $h(\rho)$
can be viewed as parameterising general coordinate transformations
of the radial variable, and its equation of motion yields the
Hamiltonian constraint $H=0$. Having obtained this equation, we can
impose $h(\rho)=1$ as a coordinate gauge condition. In this case
$H$, which must vanish, is given by
\begin{eqnarray}
H&=&\fft{a'b'}{ab} + \fft{b'^2}{2b^2} - 6 ab^2 - 2 k a b \cr &&
-\fft{\alpha}{4a^5b^6}\Big(10a^4a'^4 + 20 ab^3 a'^3b' + 22a^2 b^2
a'^2 b'^2 + 36 a^3 b a' b'^3 + 47 a^4b'^4\Big)\cr 
&& + \fft{\alpha}{a^4b^5}\Big(b(ba'+ a b')(3ba'+2ab)a'' + a(b a' + 3
a b')(b a' + 4 a b') b''\Big)\cr && +
\fft{\alpha}{a^3b^4}\Big(b^2a''^2 + 2 a b a'' b'' + 3a^2 b''^2\Big)
-\fft{2\alpha k(k a b^3 + b'^2)}{b^3}\cr
&&-\fft{\beta}{4a^5b^6}\Big(20b^4 a'^4 + 64 a b^3 a'^3 b' + 72
a^2b^2 a'^2 b'^2 + 124 a^3 b a' b'^3 + 125a^4b'^4\Big)\cr
&&+\fft{2\beta}{a^4b^5}\Big(3b(b^2a'^2+3aba'b'+a^2b'^2)a'' +
a(b^2a'^2+13ab a' b' + 16a^2 b'^2)b''\Big)\cr
&&+\fft{\beta(ba''+2ab'')^2}{a^3b^4}-\fft{2\beta(ba'+2ab')(ba'''+2ab''')
}{a^3b^4}- \fft{2\beta k(2k a b^3 + 3 b'^2)}{b^3}\,,\label{abham}
\end{eqnarray}
where a prime here denotes a derivative with respect to $\rho$.  We
may then also set $h=1$ in the effective Lagrangian, so that the
remaining equations, for $a$ and $b$, can be obtained from
\begin{eqnarray}
{\cal L}&=&\fft{a'b'}{a b} + \fft{b'^2}{2b^2} + 6 ab^2 + 2 k ab\cr
&& + \fft{\alpha}{4a^5b^6} \Big(2b^4a'^4 + 5ab^3 a'^3b' + 10 a^2 b^2
a'^2 b'^2 +12ab^3a'b'^3+ 11a^4b'4\cr &&-2ab(2b^3a'^2a'' + 2ab^2 a'b'
a'' + 3a^2bb'^2 a'' +2ab^2a'^2b'' + 4a^2ba'b'b'' + 8a^3b'^2b'')\cr
&&+2a^2b^2(b^2a''^2 + 2 a b a'' b'' +3a2b''^2)\Big) +
\fft{2k\,\alpha}{b^3}(kab^3 + b'^2 - b b'')\cr
&&+\fft{\beta}{4a^5b^6}\Big(4ka^3b^3 + 2b^2a'^2 + 2aba'b' +
5 a^2 b'^2 -2ab(ba'' + 2ab'')\Big)^2\,.\label{ablag}
\end{eqnarray}

   For the $k=0$ case, corresponding to a black brane solution,
the Lagrangian (\ref{ablag}) and Hamiltonian (\ref{abham}) have a
global scaling symmetry with
\begin{equation}
a\rightarrow \xi^2\,a\,,\qquad b\rightarrow \xi^{-1}\, b\,.
\end{equation}
This enables us to derive a conserved Noether charge, $\lambda$.
Having done this, it is more convenient now to revert to the
original radial coordinate $r$ and the metric functions $a$ and $f$ in
(\ref{bhansatz1}).  The Noether charge is then given by
\begin{eqnarray}
\lambda &=& \fft{\sqrt{f}}{ra^{5/2}}\Big[2r^2a^2(ra'-2a)
-\alpha\Big( 8a^3 f - 2 r a^2 f a' - 4r^2 a f a'^2 + 3r^3 f a'^3 +
6r a^3 f'\cr &&- 3r^3 a a'^2 f' + 3r^2a (2af-2rfa'+raf')a'' +
r^2a^2(ra'-2a) +2r^3a^2ff'''\Big)\cr &&+2\beta(ra'-2a)(4a^2f +
4rafa'-r^2fa'^2+4ra^2f'+r^2aa'f'+2r^2afa'')\Big]\,. \label{conserved}
\end{eqnarray}
It should be emphasized that this quantity is conserved only for the
case $k=0$.  The analogous Noether charge was studied in \cite{peet}
for Lifshitz black holes (with $T^2$ horizon topology) in Einstein
gravity coupled to a massive vector field. It was shown \cite{peet2}
that it is related to the energy of the black branes:
\begin{equation}
E=-\fft{\lambda\omega_2}{16\pi\,(z+2)}=\fft{2}{(z+2)} T\,S\,,\label{Elambda0}
\end{equation}
where $T$ and $S$ are the temperature and the entropy of
the black brane, and $\omega_2$ is its area.

We can test this formula with the $k=0$ Schwarzschild-AdS black
holes, corresponding to $a=f=r^2 - r_+^3/r$ in the metric ansatz
(\ref{bhansatz1}). This solution exists for all values of the
$\alpha$ and $\beta$ parameters. The temperature and the entropy are given by
\begin{equation}
T=\fft{3r_+}{4\pi}\,,\qquad S=\ft14\omega_2 r_+^2[1 - 6(\alpha + 4\beta)]\,.
\end{equation}
Note that the Gauss-Bonnet term does not contribute to the entropy in
this case. The energy is given by
\begin{equation}
E=\fft{1}{8\pi} [1 - 6(\alpha + 4\beta)] r_+^3 = \ft23 TS\,.
\end{equation}
The Noether charge $\lambda$ in this case is given by
\begin{equation}
\lambda=-6 [1 - 6 (\alpha + 4\beta)] r_+^3\,.
\end{equation}
Thus, we find that the relation (\ref{Elambda0}) holds for this
$z=1$ case.  In general we find that the second equality in
(\ref{Elambda0}) always holds, whilst the definition of energy in
terms of the Noether charge does not apply for solutions of
higher-derivative gravity when massive spin-2 modes are excited.

    For the case of Einstein-Weyl gravity, i.e. when
$\beta=-\ft13\alpha$, the Noether charge (\ref{conserved})
simplifies considerably, and becomes
\begin{eqnarray}
\lambda&=&\fft{1}{\sqrt{a^3f}\,(ra'-2a)}
\Big(2ra(18ra^2-10a^2f-2rafa'-r^2fa'^2)\cr
&&-\alpha(4ra-fa'-af')(36r^2a^2-8a^2f - rafa' - 2r^2 fa'^2-9ra^2f')
\Big)\,.\label{lambdaew}
\end{eqnarray}

\section{AdS and Lifshitz Black Holes in Conformal Gravity}

In this section, we focus on the special case of conformal gravity,
i.e. the limiting case of Einstein-Weyl gravity when $\alpha$ goes
to infinity.  The equations of motion are given by
\begin{equation}
(2\nabla^\rho \nabla^\sigma + R^{\rho\sigma})
C_{\mu\rho\sigma\nu}=0\,.
\end{equation}
Note that for the metric ansatz (\ref{bhansatz1}), the
$\alpha$-independent trace equation (\ref{trace}) does not apply in
conformal gravity.  Thus, the equation of motion is not simply
(\ref{f2peom}) with $\alpha$ sent to $\infty$.

\subsection{AdS black holes}

The most general spherically-symmetric solution in conformal gravity
was found in \cite{Riegert,Klemm,Cognola}. (See also,
\cite{mk,kks}.) The solution can easily be generalized to the other
horizon topologies $T^2$ and $H^2$.  The solution for all three
cases is given by
\begin{equation}
ds^2=-f dt^2 + \fft{dr^2}{f} + r^2 d\Omega_{2,k}^2\,,\qquad f=b r^2
+ \fft{c^2-k^2}{3d}\, r + c + \fft{d}{r}\,,\label{conformalbh}
\end{equation}
where $b$, $c$ and $d$ are arbitrary constants. The
coefficients of $r^2$ and $1/r$ are related to the excitations of the
massless graviton, while the coefficient of $r$ and the constant
$c$ are related to
the massive spin-2 mode. If $c$ is chosen so that $c=k$, the
solution reduces to the usual AdS black hole for each of the cases
$k=1$, $k=-1$ and $k=0$.  Of course, since the equations of motion
for conformal gravity leave an overall conformal factor
undetermined, it follows that $d\tilde s^2 = \Omega^2\, ds^2$ is
also a spherically-symmetric static solution, where $ds^2$ is given
by (\ref{conformalbh}) and $\Omega$ is an arbitrary function of $r$.

In fact, the solution (\ref{conformalbh}) can easily be derived by
starting from the Schwarzschild-AdS solution
\be d\tilde s^2= -\Big(k -\ft13\Lambda \rho^2 -\fft{2M}{\rho}\Big)\,
dt^2 + \Big(k -\ft13\Lambda \rho^2 -\fft{2M}{\rho}\Big)^{-1}\,
d\rho^2 + \rho^2\,d\Omega_{2,k}^2\,,\label{schadsbh}
\ee
noting that not only this, but also $ds^2=\Omega(\rho)^{-2}\,
d\tilde s^2$, is therefore a solution of conformal gravity, and then
defining a new radial coordinate via $r=\rho\, \Omega(\rho)^{-1}$.
Requiring that the resulting metric have the form $ds^2=-h dt^2 +
h^{-1} dr^2 + r^2 d\Omega_{2,k}^2$ implies that $\Omega=1 + q \rho$
where $q$ is an arbitrary constant, and hence $r=\rho/(1+q\rho)$.
The function $h$ is therefore given by
\be h= (2M q^3 + k q^2-\ft13\Lambda) r^2 -2q(k+3Mq)r + (k+ 6M q)
-\fft{2M}{r}\,,
\ee
which precisely reproduces the function $f$ in (\ref{conformalbh})
with
\be
b= 2M q^3 + k q^2-\ft13\Lambda\,,\qquad
c= k+ 6Mq\,,\qquad d= -2M\,.
\ee

The fact that the solution (\ref{conformalbh}) is related to the
usual Schwarzschild-AdS black hole does not imply that these
solutions are completely equivalent.   The scaling of the
Schwarzschild-AdS metric leaves the thermodynamic properties of the
black hole unchanged only if the conformal factor is finite and
non-vanishing in the regions between the horizon and asymptotic
infinity. However, the conformal factor $\Omega=1+q\rho$ that
relates (\ref{conformalbh}) to the usual Schwarzschild-AdS black
hole metric is in fact singular at $\rho=\infty$, and so it alters
the global structure. In turn, this affects the thermodynamic
properties, as we shall discuss below.

\subsubsection{Thermodynamics of AdS black holes in conformal gravity}

We begin by reviewing the thermodynamic properties of the standard
Schwarzschild-AdS black hole (\ref{schadsbh}) in the context of
conformal gravity. Letting $\rho_+$ denote the radius of the
outer horizon, we can solve for $M$ to get
\begin{equation}
M=\ft16 \rho_+ (3k - \Lambda \rho_+^2)\,.
\end{equation}
The Hawking temperature can be obtained from a calculation of the
surface gravity in the standard way. The entropy can be derived from
the Wald formula \cite{Iyer:1994ys}, giving
\begin{equation}
S=-\frac{\alpha}{8}\int C^{\mu\nu\rho\sigma}\epsilon_{\mu\nu}
\epsilon_{\rho\sigma}d\Sigma\,. \label{conformalentropy}
\end{equation}
Thus we have
\begin{equation}
T=\fft{k-\Lambda \rho_+^2}{4\pi \rho_+}\,,\qquad S=\ft16 \alpha
(3k-\Lambda\rho_+^2)\omega_2\,,\label{schadsts}
\end{equation}
where $\omega_2$ denotes the volume of $d\Omega_{2,k}^2$.

  It is worth remarking that at first sight the entropy of the black
hole in conformal gravity is not simply proportional to the area of
the horizon, but now it is given by
\begin{equation}
S=\ft12\alpha \Big(k\,\omega_2 + \ft13(-\Lambda)\,
A\Big)\,,\label{schadsentropy}
\end{equation}
where $A=\rho_+^2 \omega_2^2$ is the area of the horizon.  However
the first term in the above is a pure constant, independent of
the parameters in the solution, and can be removed
by introducing a Gauss-Bonnet term in the Lagrangian. In fact if we
use the action (\ref{action}) with $\beta=-\ft13\alpha$ and
$\gamma=0$, the first term in (\ref{schadsentropy}) vanishes and
hence the entropy is then proportional to the area of the horizon.

The free energy $F$ can be obtained from the Euclidean action $I_E$
of conformal gravity, using the relation $F=I_E\, T$. The action
converges for the black hole solution, leading to
\begin{equation}
F=-\fft{\alpha\omega_2}{32\pi} \int_{r_+}^\infty r^2 dr\,
|\hbox{Weyl}|^2 =-\fft{\alpha (3k - \Lambda \rho_+^2)^2 \omega_2}{72
\pi \rho_+}\,.\label{schadsf}
\end{equation}

The energy can be determined by integrating the first law,
$dE=T dS$, assuming that $\Lambda$ is held fixed, giving
\begin{equation}
E=\fft{\alpha \Lambda \rho_+ (-3k + \Lambda \rho_+^2)
\omega_2}{36\pi} = \fft{\alpha (-\Lambda) \omega_2}{6\pi}\, M\,.
\end{equation}
This expression for the energy can also be confirmed
independently by using either the Deser-Tekin
\cite{destek} or the AMD method
\cite{Ashtekar:1984zz,Ashtekar:1999jx,Okuyama:2005fg,Pang:2011cs}.

  In conformal gravity the cosmological constant $\Lambda$ is a
parameter of the solution, rather than of the theory, and hence we
may treat $\Lambda$ as a further thermodynamic variable, leading to the
more general thermodynamic relations
\begin{eqnarray}
&&dE = T dS + \Theta\, d\Lambda\,,\qquad F = E - T S\,,\qquad \Theta
= -\fft{\alpha \rho_+ (3k - \Lambda \rho_+^2)
\omega_2}{72\pi}\,.\label{schadsthermo}
\end{eqnarray}
Thus, treating the cosmological constant as a thermodynamic variable
does not affect the relationship between $F$ and $E$.  We can simply
start by assuming that $\Lambda$ is constant and obtain the first
law of thermodynamics.  The first law can then be straightforwardly
modified by treating $\Lambda$ as a variable, thus determining
the corresponding conjugate variable $\Theta$, whilst the other
thermodynamic quantities remain unchanged. Treating the cosmological
constant as a thermodynamic variable has been considered previously.
See, for example, \cite{Caldarelli:1999xj,Liu:2010sz,cgkp}. In
Einstein gravity, where the entropy is simply one quarter of the
horizon area, without explicit dependence on the cosmological
constant $\Lambda$, the quantity $\Theta\sim \rho_+^3$ is
proportional to the volume conjugate to the cosmological constant,
which can then be interpreted as a pressure \cite{cgkp}.  In
conformal gravity, on the other hand, the entropy has a manifest
dependence on $\Lambda$, and hence the quantity $\Theta$ given in
(\ref{schadsthermo}) is not simply proportional to the volume, but
has a linear $\rho_+$ dependence as well, for non-vanishing $k$. It
is also worth remarking that the Smarr formula  $E=2TS -2\Theta
\Lambda$ in Einstein gravity \cite{cgkp} now becomes $E=2\Theta
\Lambda$ in conformal gravity.  We shall discuss this further in
section 5.1.5.

We are now in a position to discuss the more general AdS black holes
in conformal gravity.  We shall begin by taking the cosmological
constant to be fixed,\footnote{When we refer to the ``cosmological
constant'' in the context of conformal gravity, where of course
there is no cosmological term in the action, we always mean the
cosmological constant of the asymptotic AdS space.} by setting $b=1$
in (\ref{conformalbh}), corresponding to setting the cosmological
constant of the asymptotic AdS space to be $\Lambda=-3$.  Letting
$r_+$ be the radius of the outer horizon and writing $d=-r_+ \tilde
d$, we find that
\begin{equation}
r_+^2 = -c + \fft{c^2-k^2}{3\tilde d} + \tilde d >0\,.
\end{equation}
The temperature and entropy can then be straightforwardly calculated;
they are given by
\begin{equation}
T=\fft{(3\tilde d-c)^2 - k^2}{12\pi r_+ \tilde d}\,,\qquad S=\ft16
\alpha (k+3 \tilde d-c) \omega_2\,.\label{gents}
\end{equation}
The free energy $F$ can also be obtained from the Euclidean
action, giving
\begin{equation}
F=-\fft{\alpha \omega_2 \Big((c-k)^2 - 3 (c-k) \tilde d + 3 \tilde
d^2\Big)}{24\pi r_+}\,.\label{genf}
\end{equation}
When $c=k$, the system reduces to the previous Schwarzschild-AdS
black hole with cosmological constant fixed at $\Lambda = -3$.  Thus,
the general solution with $c\ne k$
contains an additional independent parameter.

As we have remarked in the previous subsection, the general solution
can be obtained by performing a conformal transformation of the
Schwarzschild-AdS black hole, whose cosmological constant $\Lambda$
can be promoted to being a parameter of the solution. In the new
solution, we have chosen to set $b=1$.  Thus as a local solution,
our new variables $(c,d)$ are related to the $(M,\Lambda)$ variables
in the Schwarzschild-AdS solution (\ref{schadsbh}). It is natural to
ask whether the thermodynamics of the new solution are simply the
same as (\ref{schadsthermo}), but expressed in terms of new
variables.  In order to address this issue, we note that the
transformation described in subsection 5.1 amounts to
\begin{equation}
q=\fft{c-k}{6M}\,,\qquad M=\ft16 \rho_+(3k - \Lambda \rho_+^2)\,,
\end{equation}
with
\begin{equation}
\rho_+=\fft{3r_+ \tilde d}{k-c + 3\tilde d}\,,\qquad \Lambda =
\fft{(2k+c) r_+ - 3k\rho_+}{(r_+ - \rho_+) \rho_+^2}=\fft{(c-k - 3
\tilde d)^2 (c+2k - 3 \tilde d)}{9 r_+^2 \tilde d^2}\,.
\end{equation}
It is easy to see that when $c=k$, we have $\rho_+=r_+$ and $\Lambda
= -3$, as we should have expected.

It is straightforward to verify that the temperature and entropy in
(\ref{schadsts}) are indeed mapped into those in (\ref{gents}).
However, the free energy in (\ref{schadsf}) becomes
\begin{equation}
F\rightarrow \widetilde F = -\fft{\alpha (3 \tilde d-c+k)^3
\omega_2}{216 \pi r_+ \tilde d}\,,
\end{equation}
which is different from the free energy given in (\ref{genf}). The
reason for this can be easily understood as follows.  The $r$ and $\rho$
coordinates are related to each other by
\begin{equation}
r=\fft{\rho}{1 + q \rho}\,,\qquad \rho = \fft{r}{1-q r}\,.\label{rhor}
\end{equation}
The temperature and entropy are, in a sense, ``local'' properties,
evaluated on the horizon $r=r_+$ or $\rho=\rho_+$, and related by the
above equation. Since the theory is conformal, the temperature and
entropy are not modified by the conformal transformation. On the other
hand, the free energy, and hence the energy, are evaluated by an
integration over the regions $[r_+, \infty)$ of the general black
hole or $[\rho_+,\infty)$ of the Schwarzschild-AdS black hole. From
the relationship (\ref{rhor}), we find
\begin{eqnarray}
r_+\le r <\infty &\longrightarrow & (-\infty <\rho \le -\fft{1}{q})
\quad \cup \quad (\rho_+\le \rho <\infty)\,,\cr 
\rho_+ \le \rho <\infty & \longrightarrow & r_+ \le r <\fft{1}{q}
\,.
\end{eqnarray}
Thus, we see that the outer region $\rho\ge\rho_+$ of the
Schwarzschild-AdS black hole covers only part of the outer region
$r\ge r_+$ of the general black hole (\ref{genf}).  The exterior of
the general black hole maps into disconnected regions of the
Schwarzschild-AdS solution.  Thus, we see that although the
conformal transformation does not affect the location of the horizon
or the expression for $\sqrt{-g} |\hbox{Weyl}|^2$, the structure of
the asymptotic region is altered by the transformation.  Therefore,
the Euclidean actions are different for the two solutions.
Analogously, the energy of the two solutions, which are typically
evaluated at asymptotic infinity, are also different. The upshot is
that the two solutions cannot be viewed as equivalent.

     Having established that the new solutions are
globally inequivalent to the Schwarzschild-AdS black hole with
$(M,\Lambda)$ parameters, we shall now proceed to investigate the
thermodynamics of the general black holes in conformal gravity. We
should not expect the usual first law $dE=T dS$ still to be
satisfied, since the general solutions are now described by two
independent parameters, $c$ and $d$, rather than just one.  As we
shall see, it is necessary now to introduce an additional pair of
intensive and extensive thermodynamic variables, which we shall call
$\Psi$ and $\Xi$, and the first law will be modified to $dE=TdS+\Psi
d\Xi$. Once the additional parameter of the AdS black holes in
conformal gravity is turned on, by taking $c\ne k$, we find that
neither the  Deser-Tekin nor the AMD methods gives a finite result
for the mass. In appendix D, we describe a new procedure for
calculating the mass in conformal gravity.

   It is instructive first to look at the solution where the
parameter $d$ is set to zero. In the parameterisation in (\ref{conformalbh}),
this can be done by first writing $c^2-k^2= 3 \Xi d$, and then sending
$d$ to zero, giving
\begin{equation}
f=r^2 + \Xi \, r + k\,.\label{thermalvac}
\end{equation}
This solution has a curvature singularity at $r=0$, which can be
shielded by an horizon at $r=r_0$ provided that $\Xi$ is chosen so that
$\Xi^2\ge 4k$.  The temperature is given by
\begin{equation}
T_0=\fft{r_0^2-k}{4\pi r_0}\,.
\end{equation}
However, we find that the entropy and free energy both vanish,
suggesting that the energy should vanish also. Thus the solution can
be viewed as a ``thermalized vacuum.''  (This is
analogous to the Schwarzschild black hole in critical gravity, where
all thermodynamic quantities except for temperature vanish
\cite{lpcritical}.) In a Deser-Tekin or AMD calculation, this
thermalized vacuum will generate a divergence in the evaluation of
the mass, and it should be subtracted.

In fact it is easy to verify that this thermalized vacuum is locally
conformal to a de Sitter background.  To see this, we define
$d\hat s^2=\Omega^2 ds^2$, with
\be
\Omega= \fft1{\Xi r + 2k}\,,
\ee
and introduce the new radial coordinate $\rho=r\Omega$.  We then
have
\be d\hat s^2 = -\fft{\hat f}{4k} dt^2 + \fft{d\rho^2}{k\, \hat f} +
\rho^2 d\Omega_{2,k}^2\,,\label{vacads} \ee
where
\be \hat f = 1 -\ft13 \Lambda \rho^2\,,\qquad \Lambda = 3(\Xi^2-4k)\,.
\ee
The condition for the solution (\ref{thermalvac}) to have real roots
defining the horizons is $\Xi^2-4k\ge 0$, and so this means the
conformally-related metric $d\hat s^2$ in (\ref{vacads}) is de
Sitter spacetime, with positive cosmological constant.
The horizon in the metric with $f$ given by (\ref{thermalvac})
maps into the de Sitter horizon in (\ref{vacads}).

    From appendix D, if we take the conserved quantity $Q$ to furnish a
definition of energy, we have
\begin{equation}
E=\fft{\alpha\,\omega_2}{4\pi} (-d + m)\,,\label{genenergy}
\end{equation}
where
\begin{equation}
m\equiv \fft{(c-k)(c^2-k^2)}{18d}\,.\label{mdef}
\end{equation}
Note that when $c=k$, it reproduces the energy for the
Schwarzschild-AdS black hole.  When $d=0$, it is necessary that
$c\rightarrow k$ with $\Xi=(c^2-k^2)/(3d)$ held fixed.  In this
limit, the quantity $m$ vanishes, and hence we see that
the thermalized vacuum indeed has zero energy.

   It turns out that with this definition of energy for the general
AdS black holes in conformal gravity we have
\begin{equation}
F=E - TS\,.
\end{equation}
We find that, as mentioned earlier, the standard first law
$dE=TdS$ is not satisfied for the general AdS black holes,
since the solutions are characterised by two independent parameters.
If we first consider the situation where the quantity
$\Xi=(c^2-k^2)/(3d)$ is held fixed, then the first law $dE=TdS$
does hold.  This corresponds to allowing only variations that keep
the thermalized vacuum fixed.  If instead we allow general
variations of the two parameters in the solution, by allowing
$\Xi$ to vary also, then we find that we should add another term in
the first law, which now becomes
\begin{equation}
dE=T dS + \Psi d\Xi\,,\qquad
\Psi=\fft{\alpha \omega_2 (c-k)}{24\pi}\,.\label{newfl}
\end{equation}
The quantity $\Psi$ here is a new thermodynamic variable conjugate to
$\Xi$, which is determined from the requirement of
integrability of the generalized first law.

\subsubsection{The Noether charge of the $k=0$ solution}

As discussed in section 4.2, for $k=0$, the system has an additional
conserved Noether charge.  For conformal gravity, the Noether charge
for the ansatz (\ref{bhansatz1}) is given by
\begin{eqnarray}
\lambda &=& \fft{\alpha}{12ra^2\sqrt{a f} (ra' - 2a)}(4a^2 f-10 r a
f a'+ 7r^2f a'^2 + 6ra^2 f' - 3r^2a a' f' - 6r^2 a f a'')\cr
&&\qquad\times (4a^2 f - 2a f r a' - f r^2 a'^2 -2a^2 r f' + a
r^2a'f' + 2 a fr^2a'')\,.\label{lambdac}
\end{eqnarray}
Thus for the black hole (\ref{conformalbh}) with $k=0$, we have
\begin{equation}
\lambda = \fft{4\alpha(27d^2-c^3)}{9d}\,.
\end{equation}
For the Schwarzschild-AdS black hole (\ref{schadsbh}), we have
\begin{equation}
\lambda = 8\Lambda \alpha M\,.
\end{equation}
Note that in both cases we have $\lambda\omega_2 = -32\pi T S$. In other
words, the second equality of (\ref{Elambda0}) always holds. Indeed,
these two Noether charges for the general and the Schwarzschild-AdS
solutions can map to each other by the conformal transformation
discussed in the previous subsection. Let us define $\widetilde E$
as
\begin{equation}
\widetilde E = -\fft{\lambda\, \omega_2}{48\pi}\,.
\end{equation}
For the Schwarzschild-AdS black brane, $\widetilde E$ is precisely
the mass of the solution.  It follows from the argument presented in
the previous subsection that $\widetilde E$ cannot be the energy of
the more general solution that has an additional parameter $c$. Thus
now we have two conserved quantities; one is the true energy $E$
given in (\ref{genenergy}) and the other is $\widetilde E$.  The
difference is
\begin{equation}
E-\widetilde E = \fft{c^3}{216 \pi d}= \fft{m}{12\pi}\,,
\end{equation}
where $m$ is given in (\ref{mdef}).

\subsubsection{Conformal boundary term}

It is possible to write
a conformally-invariant boundary term in four dimensions. Thus for
completeness, this boundary term should be included in conformal
gravity. The conformal boundary term is given by
\begin{equation}
I_c=\eta \alpha \int d^3x \sqrt{-\tilde g} C^{\mu\nu\rho\sigma}
n_\mu n_\rho \nabla_\nu n_\sigma\,,\label{conformalboundary}
\end{equation}
where $n_\mu$ is the unit outward normal to the boundary, and $\eta$
is an arbitrary pure numerical constant. This boundary term does not
contribute to the equations of motion, and so it has no effect on
the local solutions, but it can contribute to the thermodynamics.
For example, it yields a non-trivial contribution to the Euclidean
action, implying that the free energy is now modified, and is given
by
\begin{equation}
F=-\fft{\alpha\omega_2[(c-k)^2 - 3(c-k)\tilde d + 3 \tilde
d^2]}{24\pi r_+} - \fft{\eta\alpha\, \omega_2\,m}{16\pi}\,,
\end{equation}
where $m$ is given by (\ref{mdef}).   It is of interest
to note that the contribution of the conformal boundary term to the
free energy is of the same form as the $m$ term appearing in the
expression (\ref{genenergy}) for the energy
in conformal gravity without the boundary term.

\subsubsection{Extremal limit}

Since the general AdS black hole (\ref{conformalbh}) has the
parameter $c-k$ in addition to the usual $d$ parameter of the
Schwarzschild-AdS black hole, it is possible to find an extremal limit
for which the temperature vanishes and the near-horizon geometry has an
AdS$_2$ factor.  For both $k=\pm 1$, the extremal solution takes the
same form, given by
\begin{equation}
f=\fft{(r-r_+)^2 (r r_+ - r_+^2 + 1)}{r r_+}\,.
\end{equation}
For $k=1$, the near horizon geometry is AdS$_2\times S^2$, with
vanishing temperature and entropy. The energy, free energy and
$\Psi$ are given by
\begin{equation}
E=F=-\fft{\alpha (r_+^2-1)^2\omega_2}{8\pi r_+}\,,\qquad
\Psi=\fft{\alpha \omega_2 (r_+^2-1)}{8\pi}\,,
\end{equation}
which all vanish for $r_+=1$. Thus the $r_+=1$ solution may also be
stable vacuum of the theory. For $k=0$, it turns out that there is
no extremal limit, since $f(r)$ has either a single root or a triple root.

\subsubsection{Thermodynamics with varying $\Lambda$}

Finally we consider the thermodynamics of the general AdS black
holes in conformal gravity when $\Lambda$, the cosmological constant
of the asymptotically AdS region, is treated as a thermodynamic
variable also.  This is natural in
conformal gravity since the cosmological constant arises as a
parameter of the solution rather than as a fixed parameter of the
theory. The solution is given by
\begin{equation}
f=-\ft13\Lambda r^2 + \Xi\, r + c + \fft{d}{r}\,, \qquad {\rm
with}\qquad 3\Xi\, d = c^2 - k^2\,.\label{solandcons}
\end{equation}
Letting $r_+$ be the radius of the outer horizon,
and defining $d=-r_+ \tilde d$, we have
\begin{eqnarray}
&& T=\fft{(3\tilde d-c)^2 - k^2}{12\pi r_+ \tilde d}\,,\qquad
S=\ft16 \alpha \omega_2 (k+3 \tilde d-c) \,,\cr 
&&\Psi=\fft{\alpha \omega_2 (c-k)}{24\pi}\,,\qquad
\Theta= \fft{\alpha \omega_2 d}{24\pi}\,,\cr 
&&F=-\fft{\alpha \omega_2 \Big((c-k)^2 - 3 (c-k) \tilde d + 3 \tilde
d^2\Big)}{24\pi r_+}\,,\qquad E= 2\Theta\, \Lambda + \Psi\, \Xi \,.
\label{fullmonty}
\end{eqnarray}
These thermodynamic quantities satisfy the relations
\begin{equation}
dE=T dS + \Theta\, d\Lambda + \Psi\, d\Xi\,,\qquad F=E - T\, S\,.
\label{genfirst}
\end{equation}
Note that the last equation in (\ref{fullmonty}) is the
Smarr formula for the general black holes in conformal gravity. Its
rather unusual form can be understood by considering the following scaling
argument.  Since the parameter $\alpha$ has  dimensions of
length-squared, $L^2$, and it is treated as a fixed parameter of the
theory (which may be set, without loss of generality, to $\alpha=1$),
it follows that the effective scaling dimensions for the thermodynamic
quantities are given by
\begin{equation}
[E]=\fft{1}{L}\,,\quad [T]=\fft{1}{L}\,,\quad [S]=1\,,\quad
[\Theta]=L\,,\quad [\Lambda]=\fft{1}{L^2}\,,\quad [\Psi]=1\,,\quad
[\Xi]=\fft{1}{L}\,.
\end{equation}
Thus if $E$ is viewed as a function of $S$, $\Lambda$ and $\Xi$,
namely $E=h(S,\Lambda,\Xi)$, then under scaling we shall have
$E= \mu h(S,\mu^{-2}\Lambda, \mu^{-1} \Xi)$.  Differentiating with
respect to $\mu$, setting $\mu=1$, and using the first law in (\ref{genfirst})
then gives the Smarr relation $E=2\Theta\Lambda + \Psi\Xi$ we found in
(\ref{fullmonty}).\footnote{A Smarr formula with more conventional coefficients
would arise if we were to view the
coupling constant $\alpha$ as another thermodynamic variable, so that the
thermodynamic quantities would all have their standard ``engineering''
scaling dimensions. We would then have a generalised first law
$dE=TdS+\Theta d\Lambda+ \Psi d\Xi + \sigma d\alpha$, where $\sigma$ is
a new thermodynamic variable conjugate to $\alpha$.  The Smarr formula
would then be $E=2TS -2\Theta\Lambda  -\Psi\Xi +2\sigma\, \alpha$.
However, since $\alpha$ is an overall parameter in conformal gravity,
including it as an additional variable represents an over-parameterisation of
the system.  This is reflected in the fact that there is then
a 1-parameter family of possible Smarr relations, with $E=\lambda TS +
2(1-\lambda) \Theta\Lambda + (1-\lambda) \Psi\Xi + \lambda\sigma\alpha$, where
$\lambda$ is an arbitrary constant.}

   The entropy of the general black hole can be decomposed as
\begin{equation}
S=\ft12 \alpha w_2\, k + \ft16\alpha\, (-\Lambda) A + 8\pi \Psi +
\ft12\alpha \omega_2 \Xi r_+\,,
\end{equation}
where $A=r_+^2 \omega_2$ is the area of the horizon, and the first
pure numerical term is the contribution from the Gauss-Bonnet term
in the action (\ref{action}) with $\beta=-\ft13\alpha $ and
$\gamma=\ft12\alpha$.

   We see from the constraint (\ref{solandcons}) on the parameters
that in the limit $c\rightarrow k$, we can either set $d=0$ with
$\Xi$ fixed, or set $\Xi=0$ with $d$ fixed. The former leads to the
thermalized vacuum (\ref{thermalvac}) and the latter leads to the
Schwarzschild-AdS black hole.

\subsection{$z=4$ Lifshitz black holes}

We find that conformal gravity admits static asymptotically-Lifshitz
black hole solutions also, both for $z=4$ and $z=0$.  We shall first
discuss the case with $z=4$.  The solution is given by
\begin{equation}
ds^2 = -r^8 f dt^2+ \fft{4 dr^2}{r^2f} + r^2
d\Omega_{2,k}^2\,,\qquad f=1 + \fft{c}{r^2} + \fft{c^2-k^2}{3r^4} +
\fft{d}{r^6}\,.\label{conflifsbh}
\end{equation}
This solution for Lifshitz black holes is locally
equivalent to the AdS black hole solution (\ref{conformalbh}) up to
an overall conformal factor.  Specifically, it can be seen that the
metric $d\hat s^2 = \Omega^2 ds^2$ with
\be
\Omega = \fft{q}{r(c+ 3 r^2-k)}\,, \label{z4conf}
\ee
becomes, after transforming to the new radial coordinate $\rho= r\Omega$,
\be
d\hat s^2= -\ft19 q^2\, \hat f dt^2 + \hat f^{-1}\, d\rho^2 +
   \rho^2\, d\Omega_{2,k}^2\,,\label{z4ads}
\ee
where
\be
\hat f = k + \fft{q}{3\rho} -\ft13\Lambda\, \rho^2\,,\qquad
\Lambda= \fft{(c^3-27d-3ck^2+2k^3)}{q^2}\,.
\ee
The conformal factor (\ref{z4conf}) is non-singular on the horizon
$r=r_+$ of the Lifshitz black hole (except, as we shall see below, in the
case of the $k=1$ extremal limit), and the horizon is mapped to
that of the conformally-related (A)dS black hole (\ref{z4ads}).
However, since the conformal factor becomes singular at $r=\infty$,
the asymptotic regions, and hence the global structure, are very
different for the two metrics.

    The equation $f(r)=0$ determines the locations
of the horizons. This yields a cubic equation for $r^2$,
which will have either three real roots or one real
root, according to whether the discriminant
\be
\Delta= -\fft1{27}(c^3-27d-3ck^2-2k^3)(c^3-27d-3ck^2+2k^3)\,,
\ee
is positive or negative.  In particular, in the case that $\Delta>0$, the
cosmological constant of the conformally-related metric (\ref{z4ads})
will be positive, and it describes a de Sitter black hole.

Using $r_+$ as usual to denote the radius of the outer horizon of the
Lifshitz black hole, we have
\begin{equation}
d=-\ft13 r_+^2(c^2-k^2+3c r_+^2 +3r_+^4)\,.
\end{equation}
We find that the temperature and the entropy are given by
\begin{equation}
T=\fft{(c+3r_+^2-k)(c+3r_+^2 +k)}{12\pi}\,,\qquad
S=-\ft16\alpha\omega_2(c+3r_+^2-k)\,.\label{z4TS}
\end{equation}
The above expressions suggest that
the constant $c$ might be spurious, since it always arises in the
combination $c+3r_+^2$. Indeed we can, locally, remove it by first making the
coordinate transformation $r^2=\tilde r^2 -c/3$, and then scaling
the metric by the factor $\tilde r^2/r^2$.  However, if $c$ is negative,
this transformation can be singular, if $c + 3 r_+^2 <0$, and so we cannot
simply use the above transformation to set $c=0$.  Indeed, one can see
from (\ref{z4TS}) that if $c=0$
then $T$ and $S$ cannot both be positive (if $\alpha>0$).
On the other hand, if $c$ is
sufficiently negative then we can arrange the parameters so that $T$ and
$S$ are both positive.

There is no obvious way to calculate the energy of an
asymptotically-Lifshitz black hole
directly (for example, the conserved charge given by (\ref{Qcf})
diverges).  We can, however, integrate the first law, $dE=TdS$,
to obtain a thermodynamic definition of the energy, up to an
undetermined additive constant.  From (\ref{z4TS})  we find
\begin{equation}
E=-\fft{\alpha \omega_2 (c+3r_+^2-k)^2 (c+3 r_+^2 + 2k)}{216\pi}\,,\label{energy1}
\end{equation}
where we have made a choice for the additive constant that is convenient
for the cases $k=1$ or $k=0$.  The energy definition for $k=-1$ will be 
given presently.
The Euclidean action for the $z=4$ Lifshitz black hole diverges for
large $r$.    We can instead define the free energy from the
thermodynamic relation $F=E-TS$, yielding
\begin{equation}
F=\fft{\alpha \omega_2 (c+3 r_+^2 -k)^2(2c+6 r_+^2+k)}{216\pi}\,.
\end{equation}

  We now examine the three cases $k=0$, 1 or $-1$ in more detail.
For $k=0$, there exists the Noether charge (\ref{lambdac}), giving
\begin{equation}
\lambda=\ft49 \alpha(c^3-27d)\,.
\end{equation}
Therefore, we see that (\ref{Elambda0}) holds for this solution.  In
particular, we have
\begin{equation}
E=\ft13 TS\,.
\end{equation}

   The $k=0$ solution has no extremal limit, since
then if the function $f$ has a double real root then it necessarily has
a triple real root.  For
$c+3r_+^2>0$ we can set $c=0$ without loss of generality, since the
conformal factor $\tilde r^2/r^2$ is
non-singular, as discussed earlier.  In this case, the positivity of
both the entropy and energy would require that $\alpha <0$.  On the other hand,
when we have $c+3r_+^2<0$, the constant $c$ cannot be set to zero, since
now the conformal factor $\tilde r^2/r^2$ runs from a negative value to
1 when $r$ goes from the horizon to infinity.  The positivity of both
the entropy and energy now requires that $\alpha >0$. When $c+3r_+^2=0$,
which would be the extremal limit for the $k=0$ black holes,
the solution instead has a naked singularity at $r=r_+$.  Thus for a given
$\alpha$, only one of the two branches ($c+3r_+^2>0$ or $c+3r_+^2<0$) is
well defined, since the entropy of one branch is positive at the price
that in the other branch it is negative.

For $k=1$, then again if $c+3r_+^2>0$ we can set the parameter 
$c=0$ without
loss of generality.  In this case, the solution has an extremal limit
with $r_+^2=1/3$, for which both the entropy and energy vanish.
For this branch of solutions, $r_+^2\ge 1/3$ and the non-negativeness
of the entropy and the energy defined by (\ref{energy1}) is guaranteed as
long as $\alpha$ is negative. If $c+3r_+^2<0$, then the parameter $c$
becomes non-trivial.  The range where $-2<c+3r_+^2<0$ in fact cannot
arise, since then the function $f$ actually has a third
positive root that is larger
than the putative largest root $r_+$, and  so $r=r_+$ is not the 
outer horizon.
If $c+3r_+^2 < -2$, the entropy and energy are
non-negative provided that $\alpha>0$. There is an extremal limit at
$c+3r_+^2=1$, but, since $c+3r_+^2>0$ we can reduce this to the
$c=0$, $r_+^2=1/3$ extremal case discussed previously.  Although the function
$f$ also has a double root, at $r=r_0$ if $c+3r_0^2=-1$, there is a 
larger positive root at $r=\sqrt{r_0^2+1}$, so this does not describe an 
extremal black hole. Note that for a given sign of $\alpha$, only one of 
the above two
branches of solutions is well defined, 
and only the branch with $c+ 3 r_+^2>0$ has an extremal limit. 
The near-horizon geometry of the
extremal limit is AdS$_2\times S^2$.

    The behavior of the metric functions is the same for the $k=-1$ 
solution as for
the $k=1$ solution.
Thus for the $c+3r_+^2\ge 0$ branch we can again set $c=0$, and
extremality occurs at $r_+^2=1/3$.  The near horizon limit of the
extremal black hole is
AdS$_2\times H^2$.  For solutions to have positive energy, we shift
the previous energy (\ref{energy1}) 
by a different additive constant, and define
\begin{equation}
\widetilde E=
-\fft{\alpha \omega_2 (c+3r_+^2-1)^2 (c+3 r_+^2 +2)}{216\pi}
  =E- \fft{\alpha\omega_2}{54\pi}\,.\label{energy2}
\end{equation}
The solution has non-negative energy and entropy
provided that $\alpha <0$.  For the $c+3r_+^2< -2$ branch,
the energy and entropy are non-negative provided that
$\alpha>0$.  The solution is extremal at $c+3r_+^2=1$, but this reduces
to the $c=0$, $r_+^2=1/3$ extremal case discussed above.
For a given $\alpha$, only one of the two branches of solutions
can be well defined.

\subsection{$z=0$ Lifshitz black holes}

We now turn our attention to the $z=0$ Lifshitz black hole, for which
the solution is given by
\begin{equation}
ds^2=-f dt^2 + \fft{4dr^2}{r^2 f} + r^2 d\Omega_{2,k}^2\,,\qquad
f=1 + \fft{c}{r^2} + \fft{c^2-k^2}{3r^4}\,.\label{z0metric}
\end{equation}
The solution has a power-law curvature singularity at $r=0$.  For
$k=0$, the singularity is naked.  The Noether charge is
given by $\lambda=-4\alpha c/3$.  Since the $k=0$ solution is not a
black hole, we cannot use this example to test the validity of
(\ref{Elambda0}).

For $k=\pm 1$, there is an horizon at the largest root of $f$, given
by
\begin{equation}
r_+^2=\ft16(\sqrt{3(4-c^2)} - 3c)\,.
\end{equation}
The requirement that $r_+^2>0$ implies that $-2\le
c<1$. It follows that the temperature and entropy are given by
\begin{equation}
T=\fft{1}{\pi (2-\fft{2\sqrt3\,c}{\sqrt{4-c^2}})}\,,\qquad
S=\ft1{12}\alpha\,\omega_2 (c+2k + \sqrt{3(4-c^2)})\,.
\end{equation}
As in the case of the $z=4$ Lifshitz black hole, here too we can
define the energy, up to an undetermined additive constant, by integrating
the first law $dE=TdS$.  Here, we find
\begin{equation}
E=\fft1{24\pi}\alpha\, \omega_2\,(c+2)\,,\label{z0energy}
\end{equation}
where we have chosen the (parameter-independent) additive constant
so that the energy is positive for $c> -2$.
Note that when $c=-2$, the solution becomes extremal, with $f$ given
by
\begin{equation}
f=\fft{(r^2-1)^2}{r^4}\,,
\end{equation}
and the energy defined in (\ref{z0energy}) vanishes.
The solution has a double root at $r=1$, with the near-horizon
geometry being AdS$_2\times S^2$ or AdS$_2\times H^2$.  For
$k=1$, the entropy vanishes in the extremal limit.

   The metric (\ref{z0metric}) is conformal to (A)dS.  Defining $d\hat s^2=
\Omega^2 ds^2$ with
\be
\Omega= \fft{q r}{c+k+r^2}\,,
\ee
we find after defining a new radial coordinate $\rho= r\Omega$ that
\be
d\hat s^2 = -\fft{q^2}{(c+k)^2}\, \hat f dt^2 + \hat f^{-1}\, d\rho^2
+ \rho^2\, d\Omega_{2,k}^2\,,
\ee
where
\be
\hat f = k + \fft{(c-k)q}{3\rho} -\ft13\Lambda \rho^2\,,\qquad
\Lambda= \fft{c+2k}{q^2}\,.
\ee
Thus the conformally-related metric describes an AdS black hole if
$c+2k<0$ and a dS black hole if $c+2k>0$.
The condition for having real roots for $r^2$ in the $z=0$ Lifshitz
black hole is that $4k^2-c^2\ge0$. In particular, if $k=+1$ then the
conformally-related metric will describe a de Sitter black hole.

As with the $z=4$ Lifshitz black hole discussed previously, here too
the conformal factor is non-singular on the horizon (except in the extremal
limit), and so the
horizon of the non-extremal $z=0$ Lifshitz black hole maps to the
horizon of the
(A)dS black hole.  Once again, however, the conformal factor becomes
singular at infinity, and the asymptotic regions of the two
conformally-related metrics are very different.

\section{AdS and Lifshitz Black Holes in Einstein-Weyl Gravity}

The existence of asymptotically AdS black holes in conformal gravity
over and above the standard Schwarzschild-AdS black holes suggests
that analogous more general solutions should exist also in
Einstein-Weyl gravity, possibly including at the critical point.
Furthermore, the existence of Lifshitz vacua in these theories and
their generalisations to Lifshitz black holes in conformal gravity
suggests that such Lifshitz black holes may also exist in
Einstein-Weyl gravity. However, no exact solutions with either type
of asymptotic behaviour have been found, beyond the usual
Schwarzschild-AdS black hole\footnote{There exists a degenerate case
with $\alpha=0$ and $8\beta\Lambda +1=0$, in which the Lagrangian is
simply $\sqrt{-g} (R-R_0)^2$. This degenerate case allows any metric
with constant scalar curvature $R_0$ to be a solution, including
some Lifshitz black holes \cite{lif1,lif2}. Since the equations of
motion in this case are reduced to a scalar rather than a tensor
equation, the system has no linear massive spin-2 excitations.}. In this
section, we establish their existence by using a numerical approach.

For $k=0$ AdS and Lifshitz black holes, by studying the horizon
expansion, we find the following general relation between the
Noether charge and the temperature and entropy:
\begin{equation}
\lambda\omega_2 = -32\pi T S\,.\label{lambdats}
\end{equation}
By examining the asymptotic behaviour at infinity, we find examples
for which the energy is given by $E=-\lambda\omega_2/(16\pi(z+2))$, and
hence the relation (\ref{Elambda0}) appears to hold in these cases.
However, as we have seen in conformal gravity discussed in the
previous section, the first equality of (\ref{Elambda0}) obtained in
\cite{peet} does not hold in general in higher-derivative gravity,
when massive spin-2 hair is involved.

\subsection{Horizon expansion}

The equations of motion for Einstein-Weyl gravity which follow from
(\ref{einsteinweyleom}) or from (\ref{Edef}) with $\beta=-\alpha/3$,
appear not to be explicitly solvable for the most general static,
spherically-symmetric solutions. We shall again consider the ansatz
(\ref{conformalbh}), and so the equations of motion for the metric
functions $a(r)$ and $f(r)$ are again given by (\ref{trace}) and
(\ref{f2peom}). As remarked previously, the Schwarzschild-AdS
metrics (\ref{schwads}) are solutions of these equations, but now we
shall have to resort to numerical methods in order to investigate
the most general static, spherically-symmetric solutions.

In order to do this, we first construct Taylor expansions for the
metric functions $a(r)$ and $f(r)$ in the vicinity of a black hole
horizon.  These will then be used to set the initial conditions for
$a$, $a'$, $f$ and $f'$ just outside the horizon, so that the
equations (\ref{trace}) and (\ref{f2peom}) for $a''$ and $f''$ can
be numerically integrated out to large distances.  It is instructive
first to look at the near-horizon expansions of $a$ and $f$ for the
Schwarzschild-AdS black hole (\ref{schwads}).  If we set
$\Lambda=-3$ as usual, and define the horizon radius $r_0$ by
$k-2m/r_0 +r_0^2=0$, then we have
\be
a=f = r^2 + k -\fft{r_0(k+r_0^2)}{r}\,,
\ee
and so the expansions are of the form
\bea
a(r)=f(r) &=& \Big(3 r_0 + \fft{k}{r_0}\Big)\, (r-r_0) -
  \fft{k}{r_0^2}\, (r-r_0)^2 +
  \Big( \fft{k}{r_0^3} + \fft{1}{r_0}\Big)\, (r-r_0)^3+\cdots
\,.\label{schwadsexp}
\eea

Since an overall constant factor in $a(r)$ can be absorbed into a
rescaling of the time coordinate, for the general solutions we can
consider a near-horizon expansion of the form
\bea
a(r) &=& (r-r_0) + a_2\, (r-r_0)^2 + a_3\, (r-r_0)^3 + a_4
\, (r-r_0)^4 +\cdots\,,
\\
f(r) &=& f_1(r-r_0) + f_2\, (r-r_0)^2 + f_3\, (r-r_0)^3
 + f_4 \, (r-r_0)^4 +\cdots\,.
\eea
(Note that these expansions are for non-extremal black
holes. The discussion for extremal black holes will be given
presently.) Substituting these expansions into (\ref{trace}) and
(\ref{f2peom}), we may then solve order by order in powers of
$(r-r_0)$, thus obtaining expressions for $a_n$ and $f_n$ with $n\ge
2$ in terms of $f_1$, $r_0$, $k$ and $\alpha$.  For example, we find
\bea
a_2 &=& \fft{3 r_0^3 + 5 f_1\, r_0^2 - 2 f_1^2\, r_0 + k\, r_0 + k\, f_1}{
f_1^2\, r_0^2} - \fft{(3 r_0^2 - f_1\, r_0 + k)}{4\alpha\, f_1^2\, r_0}\,,
\nn\\
f_2 &=& \fft{(f_1-3 r_0)(3 r_0^2 -2 f_1\, r_0 + k)}{f_1\, r_0^2}
   + \fft{3(3 r_0^2 - f_1\, r_0 + k)}{4\alpha\, f_1\, r_0}\,.
\eea
The expressions for the coefficients with higher $n$ become rapidly quite
complicated, and we shall not present them here.  They are easily
found, up to any desired order, using algebraic computing methods.

Since we have fixed the cosmological constant, by setting
$\Lambda=-3$, we see that $r_0$ and $f_1$ are non-trivial parameters
characterising the solutions for each choice of $k=0$, 1 or $-1$.
The case $f_1=3 r_0 + k/r_0$ corresponds to the Schwarzschild-AdS
solution, for which the series expansions can be found from
(\ref{schwadsexp}).

      The temperature and the entropy are given by
\begin{equation}
T=\fft{\sqrt{a_1 f_1}}{4\pi}\,,\qquad S=\ft14\omega_2 r_0\Big(r_0 +2\alpha
(f_1 - 2r_0)\Big)\,.
\end{equation}
The entropy is calculated with respect to the action of Einstein-Weyl
gravity. When $k=0$, the Noether charge is given by
\begin{equation}
\lambda=-2 \sqrt{a_1f_1} r_0 \Big(r_0-2\alpha (2 r_0-f_0)\Big)\,.
\end{equation}
It follows that the relation (\ref{lambdats}) indeed holds in
general. In the above entropy calculation, we used the action
(\ref{action}) with the Gauss-Bonnet term set to zero ($\gamma=0$).
The Gauss-Bonnet term contributes $S_{\rm GB}=\gamma\,k$, which is a
purely numerical constant, independent of the metric modulus
parameters.

     In the above consideration, the functions $a$ and $f$ have the
same single root $r=r_0$, giving rise to non-extremal black holes.
In the extremal limit, these functions have a double root, so that
the near-horizon geometry has an AdS$_2$ factor.  The Taylor
expansion is given by
\begin{eqnarray}
a(r) &=& (r-r_0)^2 + a_3\, (r-r_0)^3 + a_4 \, (r-r_0)^4
+\cdots\,,\cr 
f(r) &=& f_2\, (r-r_0)^2 + f_3\, (r-r_0)^3 + f_4 \, (r-r_0)^4
+\cdots\,.\label{extremalhor}
\end{eqnarray}
We find that the leading-order expansion of the equations of motion
when $r\rightarrow r_0$ implies that
\begin{equation}
(4\alpha -1) (3r_0^2 + k)=0\,.\label{extremalcons}
\end{equation}
Thus we see that for $k=0,1$, extremal black holes do not exist
except for $\alpha =\ft14$, on which we shall focus.  Taking
this $\alpha$ value, we find that
\begin{eqnarray}
&&a_3=-\fft{2 (4 k + 15 r_0^2)}{3 r_0 (k + 6 r_0^2)}\,,\qquad
a_4=\fft{41 k^2 + 358 k r_0^2 + 759 r_0^4}{9 r_0^2 (k + 6
r_0^2)^2}\,,\cr 
&&f_2=\fft{k + 6 r_0^2}{r_0^2}\,,\qquad f_3=-\fft{2 (2 k + 9
r_0^2)}{3 r_0^3}\,,\qquad f_4=\fft{15 k^2 + 148 k r_0^2 + 363
r_0^4}{9 r_0^4 (k + 6 r_0^2)}\,.
\end{eqnarray}
As one would have expected, the Noether charge of the $k=0$ solution
vanishes in the extremal limit.  Note that in the extremal limit,
there is only one non-trivial parameter $r_0>0$.  As we shall
discuss presently, these near-horizon geometries can extend smoothly
to the asymptotic AdS or Lifshitz infinities.  When $k=-1$, the
constraint (\ref{extremalcons}) can be solved with $r_0^2=1/3$ for
arbitrary $\alpha$.  However, the resulting solution is simply the
Schwarzschild-AdS solution whose $f$ can have a double zero for
$k=-1$.

\subsection{Asymptotic expansion}

\subsubsection{Asymptotically AdS solutions}

For asymptotically AdS solutions, the asymptotic regions behave roughly
as follows:
\begin{eqnarray}
a&\sim& r^2(1+c_0) + k - \fft{2M}{r} + c_1 r^{n+1} +
\fft{c_2}{r^n}\,,\cr f&\sim& r^2(1+c_0) + k - \fft{2M}{r} -\ft13
c_1(n-1)r^{n+1} + \fft{c_2(n+2)}{3r^n}\,,\label{adsinfty}
\end{eqnarray}
where we have parameterized $\alpha$ by
\begin{equation}
\alpha = -\fft{1}{n(n+1)}\qquad \implies\qquad n=-\ft12 \pm
\ft12\sqrt{1-\fft{4}{\alpha}}\,.
\end{equation}
Note that there are a total of four parameters in (\ref{adsinfty}),
corresponding to four excitations.  The coefficients $(c_0,M)$
correspond to the massless spin-2 modes, whilst the $(c_1,c_2)$
correspond to the massive spin-2 modes. For $0<\alpha<4$, the
constant $n$ is complex, implying that the excitation takes the form
\begin{equation}
\sqrt{r}\Big(c_1\, \cos(\ft12\sqrt{4/\alpha-1}\, \log r) +
c_2\,\sin(\ft12\sqrt{4/\alpha-1}\, \log r)\Big)\,.
\end{equation}

Let us present some explicit examples. The first is $n=-1/2$,
corresponding to $\alpha=4$.  The functions $a$ and $f$ at
asymptotic infinity are given by
\begin{eqnarray}
a&=& r^2 +m \sqrt{r}+ k -\fft{2M}{r} +\fft{5k\,m}{16r^{3/2}} -
\fft{m(m^2+48M)}{96r^{5/2}} + \cdots\,,\cr
f&=& r^2 +\ft12m\sqrt{r} + k- \fft{32M+7m^2}{r}
-\fft{15k\,m}{32r^{3/2}} -
\fft{5m(16M+m^2)}{64r^{5/2}} + \cdots\,.
\end{eqnarray}
When $k=0$, we have the Noether charge $\lambda=-27(8M+m^2)/2$.  In
this case, the usual Deser-Tekin and AMD methods of energy
calculation lead to divergent results, and hence we do not have an
independent method of calculating $E$ to verify whether the first
equality of (\ref{Elambda0}) holds.

The second example is $n=1/2$, corresponding to $\alpha=-4/3$. We
have
\begin{eqnarray}
a&=& r^2 + k +\fft{m}{r^{\fft12}} -\fft{2M}{r}
+\fft{11k\,m}{96r^{\fft52}} + \fft{m^2}{12r^3} -
\fft{Mm}{6r^{\fft72}}+\cdots\,,\cr 
f&=& r^2 + k +\fft{5m}{6r^{\fft12}} -\fft{2M}{r}
+\fft{65k\,m}{192r^{\fft52}} + \fft{25m^2}{144r^3} -
\fft{Mm}{4r^{\fft72}}+\cdots\,.
\end{eqnarray}
For $k=0$, the Noether charge is $\lambda=20M$, which is independent
of $m$. In principle, the asymptotic behavior could have the
$r^{3/2}$ series as well, but it does not appear to be the case.

The third example is $n=2$, corresponding to $\alpha=-1/6$, for which we find
\begin{eqnarray}
a&=&r^2 + k -\fft{2M}{r} + \fft{m}{r^2} -\fft{5k\,m}{21r^4} +
\fft{Mm}{3r^5}+\cdots\,,\cr 
f&=&r^2 + k -\fft{2M}{r} + \fft{4m}{3r^2}
-\fft{10k\,m}{21r^4} + \fft{5Mm}{9r^5}+\cdots\,.
\end{eqnarray}
For $k=0$, the Noether charge is $\lambda=-8M$. The energy density
can be calculated using the Deser-Tekin or AMD methods, giving
$E=M/(6\pi)$. Thus for this case, the first equality in
(\ref{Elambda0}) holds.  However, the numerical results indicate
that only the $m=0$ case, i.e. the Schwarzschild-AdS solution,
describes a black hole with an horizon.

The final example is the critical point, namely $\alpha=-1/2$,
corresponding to $n=1$.  We find that
\begin{eqnarray}
a &=& r^2 + k - \fft{2\tilde M}{r} - \fft{7k\,m}{15r^3} +
\fft{2m\,\tilde M}{3r^4} + \cdots\,,\cr 
f &=& r^2 + k - \fft{2\tilde M -2m/3}{r} - \fft{k\,m}{r^3} +
\fft{9(18\tilde M + 7m)}{18 r^4} + \cdots\,, \label{log-bh}
\end{eqnarray}
where
\begin{equation}
\tilde M=m \log r + M\,.
\end{equation}
For $k=0$, the Noether charge is $\lambda=-18m$.  In appendix C, we
derive the mass formula for this case, and we find the energy
density is $E=3m/(8\pi)$.  Thus, in this case, the first equality of
(\ref{Elambda0}) holds.

\subsubsection{Asymptotic Lifshitz behavior}

In this case, we are primarily concerned with the $k=0$ case.  We
find that the large $r$ expansion is given by
\begin{eqnarray}
a &\sim & r^{2z}\Big(1+\fft{(z^2+2)m}{z(z+2)r^{z+2}} + \tilde c_+
r^{-1-\fft12z + \fft12\Delta} + \tilde c_- r^{-1-\fft12 z -
\fft12\Delta}\Big)\,,\cr f%
&\sim & \sigma r^2\Big(1 + \fft{m}{r^{z+2}} + c_+ r^{-1-\fft12z +
\fft12\Delta} + c_- r^{-1-\fft12 z - \fft12\Delta}\Big)\,,\cr %
\tilde c_+&=&\fft{4-11z+z^2-3z^3 + (1-3z-z^2)\Delta}{2(z-1)^2(3z-1)}
c_+\,,\qquad \Delta = \sqrt{3(4-4z+3z^2)}\,,\cr \tilde
c_-&=&\fft{4-11z+z^2-3z^3 - (1-3z-z^2)\Delta}{2(z-1)^2(3z-1)} c_-\,.
\end{eqnarray}
The Noether charge is given by
\begin{equation}
\lambda=-\fft{6\sqrt6 (2-3z+z^3)m}{z^2(z-4)\sqrt{z^2+2z+3}}\,.
\end{equation}
Our numerical results suggest that there are Lifshitz-like black
holes with $S^2$ and $H^2$ topology that have the same leading-order
behavior as the above.

\subsection{Numerical analysis}

We have carried out a numerical analysis for a variety of choices
for the coefficient $\alpha$ that multiplies the Weyl-squared term
in the action. The choice of the horizon radius $r_0$ is a
non-trivial parameter, given that we have fixed the cosmological
constant ($\Lambda=-3$).  The value of the expansion coefficient
$f_1$ is also a non-trivial parameter in the solutions.  The
deviation of $f_1$ from the value $3 r_0 + k/r_0$ determines the
deviation of the black hole from the usual Schwarzschild-AdS
solution.

The Schwarzschild-AdS black hole can be thought of as a solution
where only the massless spin-2 modes are excited.  Deviating from
$f_1= 3 r_0 + k/r_0$ corresponds to setting initial conditions near
the horizon that cause the massive spin-2 modes to be excited also.
Our numerical investigations suggest that solutions of this type
exist, in the sense that the numerical routines give a reasonably
stable result with the metric functions showing no sign of runaway
behaviour, provided that the linearised spin-2 massive mode falls
off less rapidly than the spin-2 massless mode. This fall-off rate
is governed by the mass $m$ of the linearised fluctuation, and in
turn, this is related to the value of the parameter $\alpha$ in the
Lagrangian. Specifically, the condition of less rapid fall-off is
achieved if the massive mode has negative mass-squared.  Solutions
with stable behaviour appear to exist regardless of whether the
negative $m^2$ lies in the non-tachyonic region $-\ft94\le m^2<0$ or
the tachyonic region $m^2<-\ft94$.  Of course in the latter case one
would expect the solutions to exhibit time-dependent runaway
behaviour, but this will not show up with the static metric ansatz
that we are considering here.

 In terms of the constant $\alpha$ that characterises the coefficient
of Weyl-squared in the action, the condition that the massive
linearised mode have $m^2<0$ corresponds to $\alpha <-\ft12$ or
$\alpha>0$.

We find that if $\alpha$ lies in the region $-\infty <\alpha
<-\ft12$, then defining
\be
f_1= 3 r_0 + k/r_0 + \delta\,,\label{delta}
\ee
there is a range for $\delta$, with $\delta_- <\delta <\delta_+$,
for which the numerical solutions indicate the occurrence of
asymptotically AdS black holes.  The lower limit $\delta_-$ is
negative, while the upper limit $\delta_+$ is positive.  If the
value of $\delta$ is fine-tuned to be {\it equal} to $\delta_-$ or
$\delta_+$, then the asymptotic behaviour of the black hole changes
from AdS to Lifshitz.  The value of $z$ in the asymptotically
Lifshitz case is given by the larger root in (\ref{zsol}).  If the
parameter $\delta$ is chosen to lie outside the range $\delta_-\le
\delta\le \delta_+$, then the numerical analysis indicates that the
solution becomes singular.

   As an example, let us consider $\alpha = -\fft{11}{16}$, which from
(\ref{zsol}) implies that there should exist asymptotically Lifshitz
solutions with $z=2$.  Taking $k=0$ and choosing $r_0=10$, we find
that the limiting values for $\delta$  in (\ref{delta}) are
\be
\delta_-\approx -11.596956988\,,\qquad
\delta_+\approx  62.826397763\,.\label{k0range}
\ee
In our numerical routine, we set initial conditions just outside the
horizon at $r=r_0 + 0.0001$, and run out to $r=100000$.  For the
asymptotically Lifshitz black hole with $\delta=\delta_-$, we
obtained plots of $a(r)$, $f(r)$, given in Figure 1, and $a(r)/r^4$
and $f(r)/r^2$, given in Figure 2. Note that although we integrated
out to $r=100000$, we only plot the functions out to $r=100$ in
order to be able to generate more illustrative displays.   The
asymptotic value of the ratio $f(r)/r^2$ reaches about
$0.545454545452$ as $r$ approaches 100000, which is indeed close to
the expected ratio $6/11$ (see equation (\ref{sigma})).

\begin{figure}[htbp]
\centerline{\epsfxsize=2.4truein \epsffile{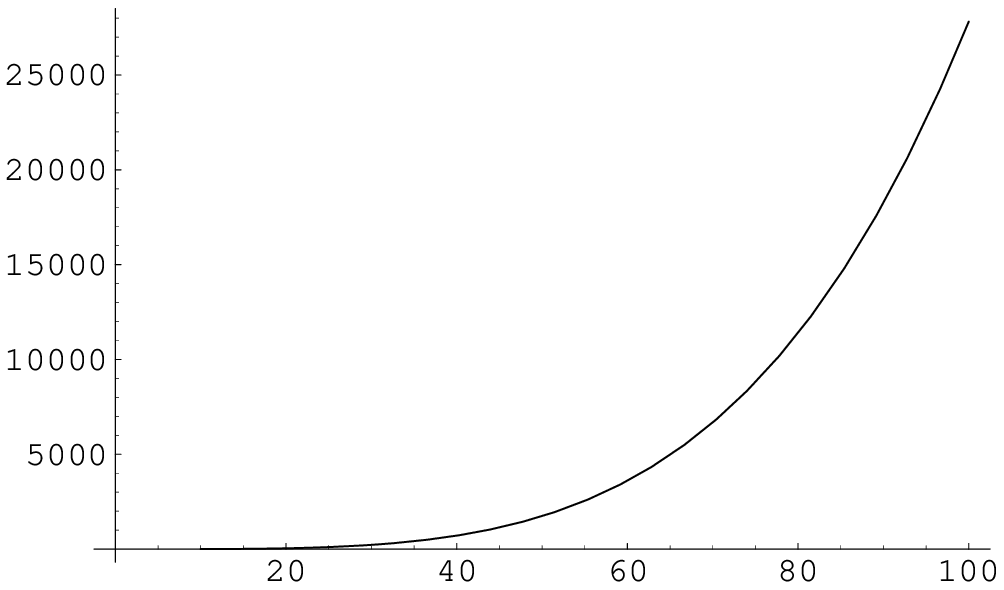} \hspace{0.25in}
\epsfxsize=2.4truein \epsffile{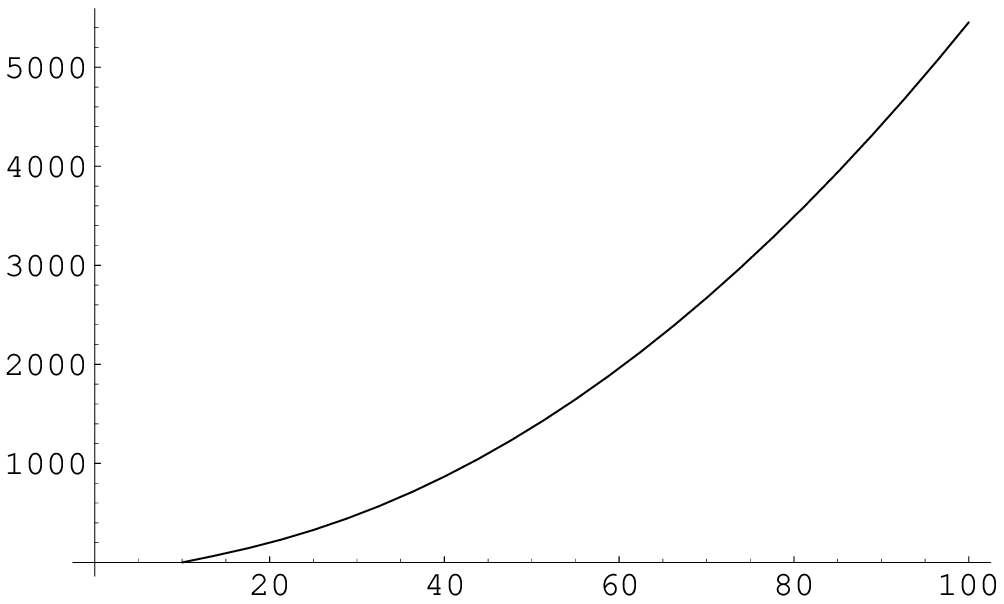} } \caption{The metric
functions $a(r)$ and $f(r)$ for the asymptotically Lifshitz black hole.}
\end{figure}

\begin{figure}[htbp]
\centerline{\epsfxsize=2.4truein \epsffile{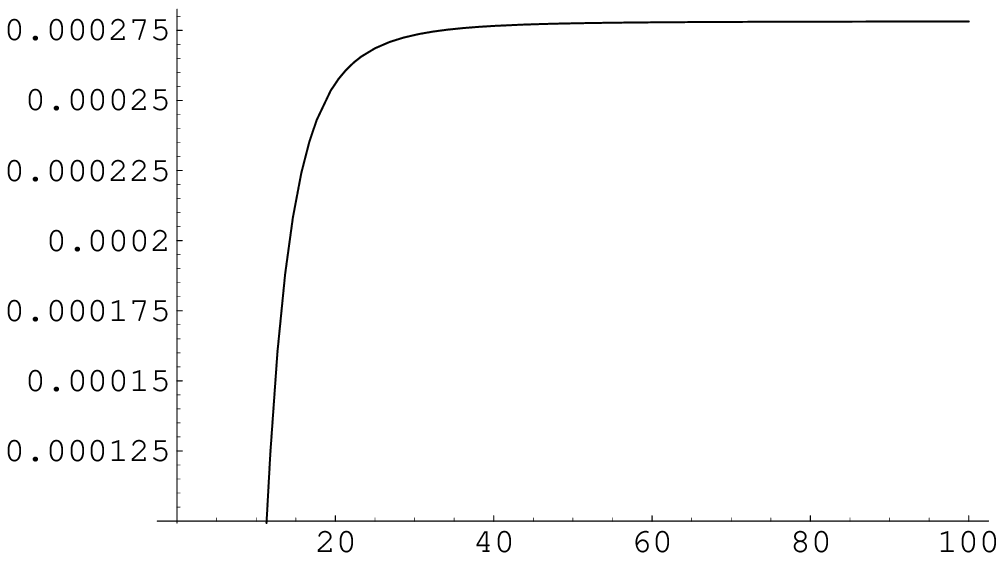}
\hspace{0.25in} \epsfxsize=2.4truein \epsffile{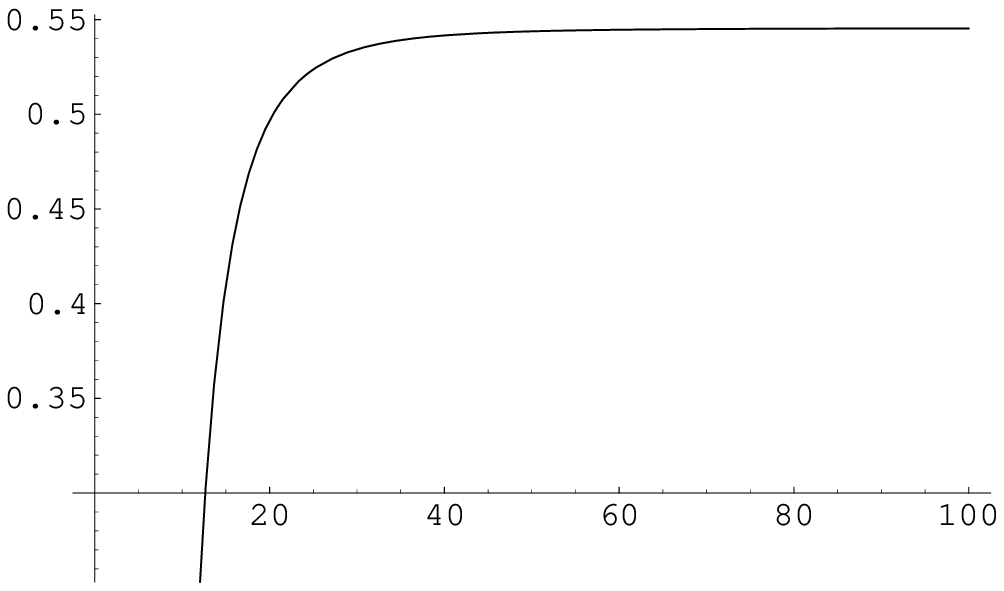} }
\caption{The asymptotic forms for $a(r)/r^4$ and $f(r)/r^2$,
illustrating the $z=2$ Lifshitz behaviour.}
\end{figure}

The solution with $\delta=\delta_+$ exhibits very similar Lifshitz
behaviour.  If we choose a value of $\delta$ that lies in between the
two Lifshitz extremes, we obtain an asymptotically AdS black hole.
Figures 3 and 4 below illustrate this, again for $\alpha=-\ft{11}{16}$,
$k=0$ and $r_0=10$, in the case that $\delta=20$.  Figure 3 shows
the functions $a$ and $f$, while Figure 4 shows the functions
$a/r^2$ and $f/r^2$.

\begin{figure}[htbp]
\centerline{\epsfxsize=2.4truein \epsffile{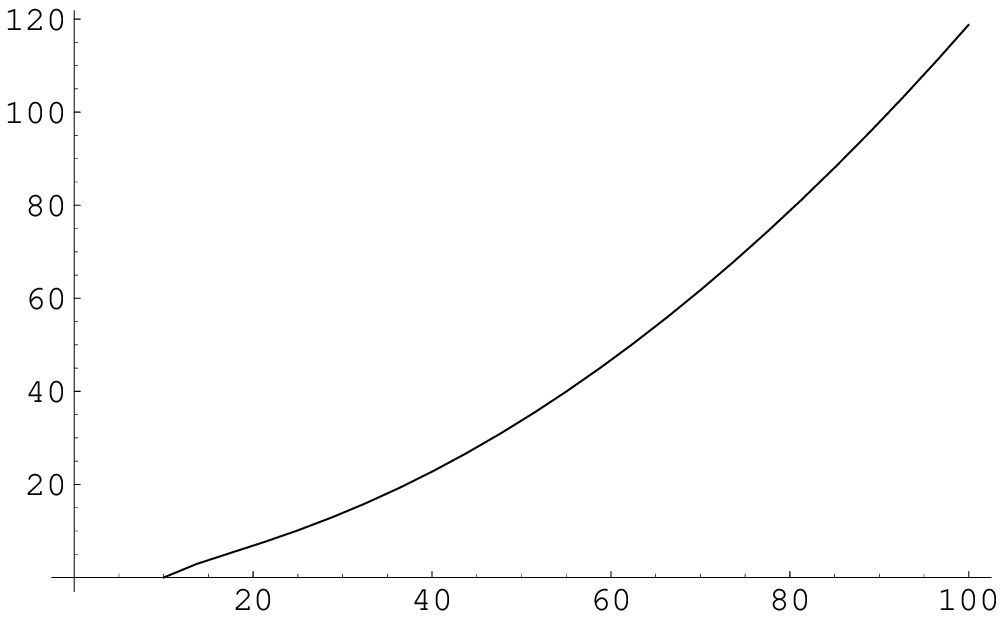} \hspace{0.25in}
\epsfxsize=2.4truein \epsffile{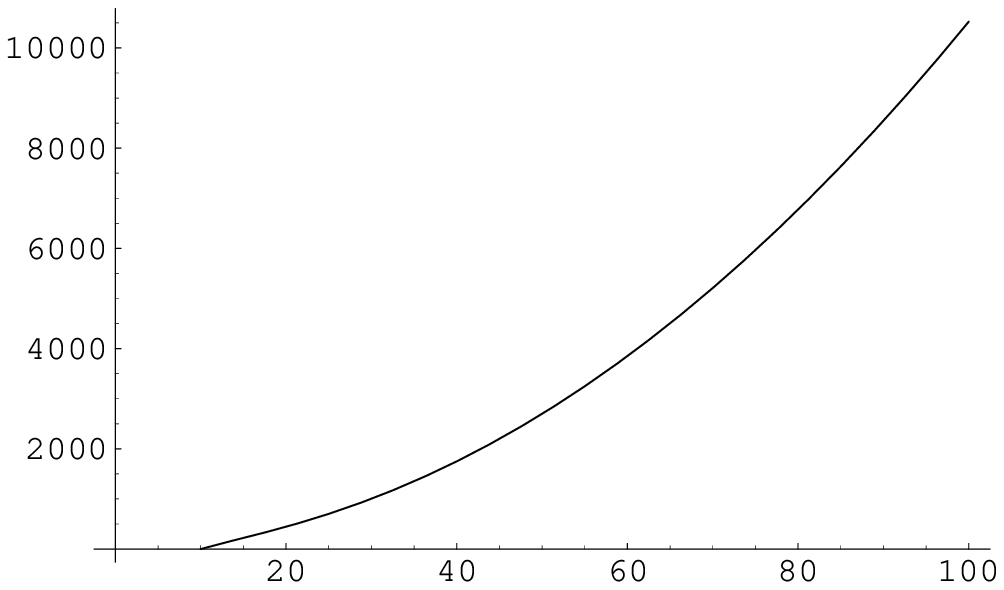} } \caption{The metric
functions $a(r)$ and $f(r)$ for the asymptotically AdS black hole.}
\end{figure}

\begin{figure}[htbp]
\centerline{\epsfxsize=2.4truein \epsffile{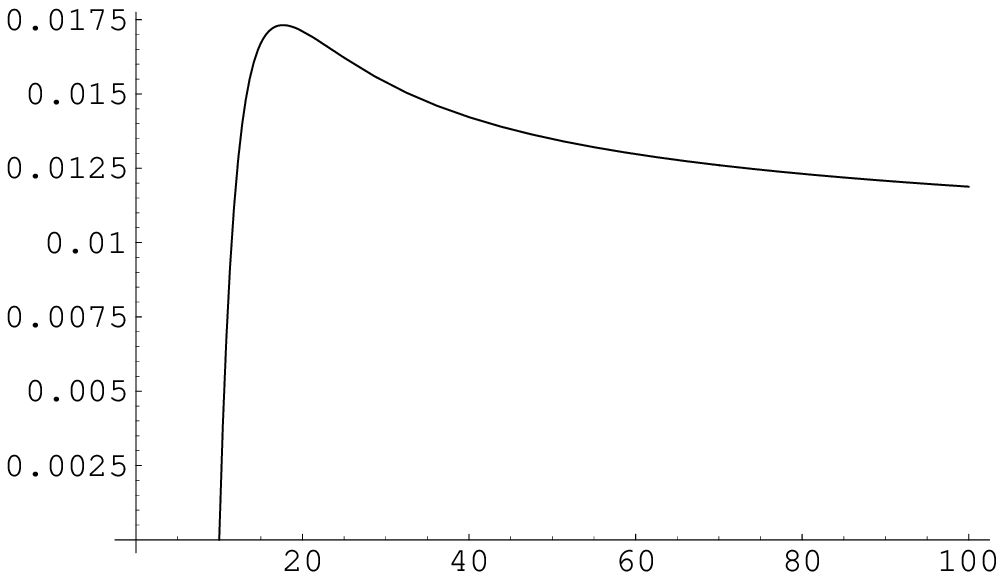}
\hspace{0.25in} \epsfxsize=2.4truein \epsffile{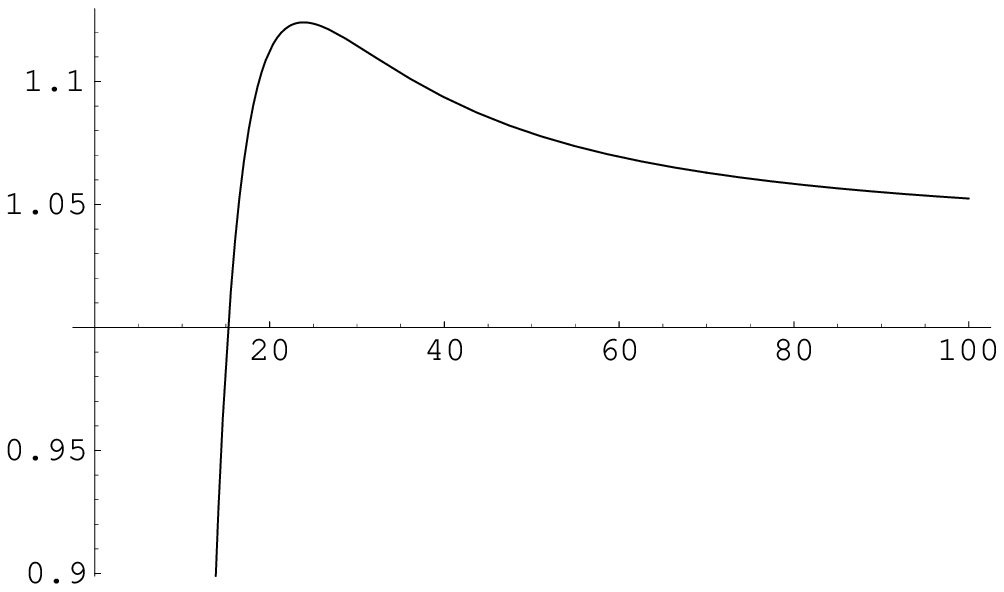} }
\caption{The asymptotic forms for $a(r)/r^2$ and $f(r)/r^2$,
illustrating the AdS behaviour.}
\end{figure}

The story is very similar for solutions with $k=1$ or $-1$.  For
example, if we consider $k=1$ solutions, again with
$\alpha=-\ft{11}{16}$ and $r_0=10$, we find that the upper and lower
limits on the range of $\delta$ in (\ref{delta}) is now
\be
\delta_-\approx -11.596956988\,,\qquad
\delta_+\approx  62.826397763\,.\label{k1range}
\ee
We find solutions exhibiting asymptotically Lifshitz type of
behaviour, again with $z=2$, if $\delta$ is taken to be either of
the extreme values. If, on the other hand, $\delta$ lies in between
the limiting values $\delta_-$ and $\delta_+$, then we find
solutions with asymptotically AdS behaviour.  The forms of the
metric functions $a$ and $f$ are qualitatively similar to those
illustrated in the $k=0$ examples above.

   When $\alpha=-\ft12$, corresponding to the case of critical gravity,
numerical analysis indicates that asymptotically AdS black hole solutions
again exist, within some range of values for the $\delta$ parameter
in (\ref{delta}).  However, $\alpha=-\ft12$ is on the borderline for
stability of the solutions, with $-\ft12 <\alpha<0$ seemingly being
unstable, and so it is not easy to extract meaningful quantitative
results in the critical case.

    We also perform the numerical analysis for the extremal case
with the parameter $r_0$, whose horizon behavior is given by
(\ref{extremalhor}).  For $S^2$ or $T^2$ topology, such a solution
exists only for $\alpha=1/4$.  For $k=1$, the numerical results
indicate that the horizon can smoothly extend to the asymptotic
AdS$_4$ infinity for all parameters $r_0>0.251976578$. When
$r_0=0.251976578$, the asymptotic behavior becomes Lifshitz-like
with exponent $z=-1/2$.  For $r_0<0.251976578$, the solution becomes
singular. For $k=0$, the horizon can extend smoothly to AdS in the
asymptotic region for any $r_0>0$.  For $k=-1$, we must have $r_0
>1/\sqrt6$. When $r_0=1/\sqrt3$, the usual Schwarzschild-AdS solution
emerges. There is no indication of Lifshitz behavior for $k=-1$.

\section{Conclusions}

In this paper, we have considered four-dimensional Einstein gravity
extended by the addition of general quadratic-curvature terms.  In
addition to the usual AdS vacuum, the theory contains Lifshitz and
Schr\"odinger vacuum solutions.  Our primary purpose was to
construct black holes obeying asymptotically AdS or Lifshitz
boundary conditions, with spherical, 2-torus or hyperbolic $H^2$
spatial symmetry. We focused on conformal gravity, with a
purely Weyl-squared action, as well as Einstein-Weyl gravity, for
which the standard Einstein action with cosmological constant is
augmented with a Weyl-squared term. The general
spherically-symmetric local solution in conformal gravity was known
previously.  It involves two nontrivial parameters, one of which is
associated with the mass of the black hole while the other, which we
call $\Xi$, may be thought of as characterising massive spin-2 hair.

   Owing to the presence of the second non-trivial parameter in the
general AdS black hole solutions, one can expect that the usual
first law of thermodynamics, $dE=TdS$, will need to be augmented by
an additional term involving a new pair of intensive and extensive
thermodynamic variables.  We studied this in detail in the case of
AdS black holes in conformal gravity, showing how the first law
becomes $dE=TdS+\Psi d\Xi$, where the variable $\Psi$, conjugate to
$\Xi$, is determined by requiring the integrability of the equation.
We also needed to find a satisfactory definition of energy for the
black holes in conformal gravity.  Its derivation, as a conserved
charge evaluated at infinity, is described in appendix D.

 In conformal gravity, the cosmological constant $\Lambda$ of the AdS
black holes is a parameter of the solution rather than a parameter
in the action; it characterises the ``AdS radius'' of the
asymptotically AdS region.  It is therefore natural to promote
$\Lambda$ to being another thermodynamic quantity that can be varied
in the first law.  We showed that this indeed gives a consistent
extension of the thermodynamic phase space.

  We then constructed Lifshitz black holes in conformal gravity, with
a temporal/spatial anisotropic scaling parameter $z=4$.  These
solutions involve only a single non-trivial parameter, and hence the
thermodynamic quantities can be easily evaluated. We showed that,
since the Lifshitz black hole has $T^2$ spatial sections, there
exists a conserved Noether charge $\lambda$.  Moreover, $\lambda$ is
related to the energy of the black hole, and to the product of
temperature and entropy (\ref{Elambda0}), in the same way as has
previously been observed in \cite{peet} for certain two-derivative
theories. However, for the more general AdS black holes (with $T^2$
spatial sections) involving massive spin-2 hair, characterised by $\Xi$,
the Noether charge no longer seems to provide a natural definition
for the energy, although the second equality of (\ref{Elambda0})
always holds.  By contrast, in the case of Schwarzschild-AdS black
holes with $T^2$ spatial sections, the relation (\ref{Elambda0})
always holds.  We also obtained Lifshitz-like black holes with $S^2$
and $H^2$ spatial sections, with Lifshitz exponent $z=4$ and 0, and
we found that the thermodynamic relations are obeyed in these cases.

    The existence of well-defined AdS and Lifshitz black holes in
conformal gravity with additional massive spin-2 hair prompted us to
seek similar solutions in Einstein-Weyl gravities.  It does not
appear to be possible to obtain closed-form expressions for such
solutions, and so we resorted to numerical integration of the
equations of motion.  The procedure is to first obtain both the
horizon and asymptotic expansions, and then use the horizon
expansion as the initial boundary conditions for numerical analysis
and compare the resulting solution for large radial values with the
asymptotic expansions.  We find that the horizon geometry involves
an extra parameter over and above that of the usual
Schwarzschild-AdS solution which is an Einstein metric.  The
numerical analysis suggests that asymptotically AdS black holes
exist within a continuous range of values for the additional
parameter.  At the boundary of this parameter region, the asymptotic
behavior changes to that of Lifshitz solutions, giving rise to
corresponding asymptotically Lifshitz black holes.  Beyond these
parameter boundaries, the solutions develop naked curvature
singularities. For a black brane with $k=0$, for which there is an
additional Noether charge, we find that the second equality in
(\ref{Elambda0}) always holds, whereas the first equality does not.
These solutions provide an interesting phase transition of the
corresponding boundary field theory from a relativistic Lorentzian
system to a non-relativistic Lifshitz system.  We further examine
the existence of extremal solutions whose near-horizon geometry has
an AdS$_2$ factor.  It turns out that non-trivial AdS extremal
solutions arise only for $\alpha=\fft14$.  In the case of $k=1$,
there exists an extremal Lifshitz-like black hole with exponent
$z=-\fft12$.

It would be interesting to explore the possibility of embedding
extended gravity within string theory, given that string theory
contains higher derivative corrections due to stringy or quantum
effects. In fact, other than some special cases for which there is
maximal supersymmetry, not much is known regarding the forms of
these higher derivative corrections. In light of the vast string
landscape, one expects that there are generic corrections, which
include the higher-order curvature terms discussed in this paper.
One might then invoke holographic techniques, in which case the
black hole solutions discussed in this paper could be used to
describe three-dimensional field theories or condensed matter
systems. For the AdS black holes, the extra parameter of the AdS
black hole solution would be mapped to a parameter in the dual field
theory associated with finite coupling corrections. The additional
global symmetry exhibited by the AdS black brane solutions would
then be associated with a particular scaling symmetry in which space
and time are rescaled differently, which is present at the conformal
fixed point as well as away from it.

\section*{Acknowledgements}

H.L.~is grateful to the George P. and Cynthia W. Mitchell Institute
for Fundamental Physics and Astronomy for hospitality during the
course of this work. The research of H.L.~is supported in part by
NSFC grant 11175269. The research of C.N.P.~is supported in part by
DOE grant DE-FG03-95ER40917. The research of J.F.V.P.~is supported
in part by NSF grant PHY-0969482 and a PSC-CUNY Award.
C.N.P.~gratefully acknowledges the hospitality of the Mitchell
Family Foundation at Cook's Branch Conservancy during the completion
of this work.

\appendix

\section{Further Solutions in Conformal Gravity}

\subsection{Asymptotically Schr\"odinger solutions}

Here we construct solutions that are asymptotic to the Schr\"odinger
solutions discussed in \cite{Balasubramanian:2008dm}.  We consider
the metric ansatz
\begin{equation}
ds^2=-r^{2z} f dt^2 + \fft{dr^2}{r^2 f} + r^2 (-2dt dx + dy^2)\,,
\end{equation}
for $z=(1,\fft12,0,-\fft12)$.  We find that the equations are
reduced to the fourth-order differential equation
\begin{eqnarray}
0&=&f'''' + \fft{(6f+18zf + 7rf')f'''}{2rf} + \fft{f''}{2r^2f^2} \Big(
4r^2ff''+ 2r^2 f'^2 + (53z + 5) r f f'\cr
&& + 4(16z^2 + 3 z - 1) f^2\Big)
+\fft{f'}{2r^3 f^2} \Big(2(3z - 1)r^2 f'^2 + (78 z^2- 11z
-9)r f f'\cr
&& + 4(28 z^2 - 10 z^2 - 7z + 1) f^2\Big) +
\fft{4z(z-1)(2z-1)(2z+1)f}{r^4}\,.
\end{eqnarray}
For $z=1$, we find a solution $f=1-M/r$, giving
\begin{equation}
ds^2=-r^{2} f dt^2 + \fft{dr^2}{r^2 f} + r^2 (-2dt dx +
dy^2)\,,\qquad f=1-\fft{M}{r}\,.
\end{equation}

\subsection{Generalized Plebanski metric}

Using the Plebanski metric ansatz \cite{pleb}, we find solutions in
conformal gravity given by
\begin{equation}
ds^2=-\fft{\Delta_x}{x^2 + y^2} (dt + y^2 d\psi)^2 +
\fft{\Delta_y}{x^2+y^2} (dt-x^2 d\psi)^2 + \fft{x^2+y^2}{\Delta_x}
dx^2 + \fft{x^2 + y^2}{\Delta_y} dy^2\,,
\end{equation}
where
\begin{eqnarray}
\Delta_x &=& c_0 + c_1 x + c_2 x^2 + c_3 x^3 + c_4 x^4\,,\cr
\Delta_y &=& c_0 + d_1 y - c_2 y^2 + \fft{c_1c_3}{d_1} y^3 + c_4 y^4\,.
\end{eqnarray}
This metric is conformal to the Plebanski-Demianski \cite{plebdem}
metric. Namely, the metric $d\hat s^2=\Omega^2 ds^2$ with
$\Omega=(1+c_3 x\,y/d_1)^{-1}$ is Einstein and satisfies $\hat
R_{\mu\nu}=\Lambda \hat g_{\mu\nu}$ with
\begin{equation}
\Lambda=-\fft{3(c_0 c_3^2+c_4 d_1^2)}{d_1^2}\,.
\end{equation}

\section{Further Solutions in Extended Gravity}

\subsection{Time-dependent metrics}

For the general quadratic action (\ref{action}) with arbitrary
$\alpha$ and $\beta$, we will consider time-dependent and spatially
flat isotropic solutions described by the metric
\be
ds^2=-dt^2+f(t)^2 \sum_{i=1}^3 dx_i^2\,.
\ee
Applying the trace condition reduces the equations to the single
third-order equation
\be
3f^2 f'^2-\Lambda f^4-6(\alpha+3\beta) \left( 3f'^4-2ff'^2
f''+f^2 f''^2-2f^2f'f'''\right) =0\,. \ee
For Einstein-Weyl gravity, $\alpha+3\beta=0$ and, for a positive
cosmological constant, the only solution is de Sitter spacetime.

We will now consider non-isotropic time-dependent solutions for
extended gravity with zero cosmological constant, described by the
Kasner metric
\be ds^2=-dt^2+\sum_{i=1}^3 t^{2p_i} dx_i^2\,. \ee
It can be shown that a metric of this form must satisfy the conditions
\be\label{conditions} \sum_{i=1}^3 p_i=\sum_{i=1}^3 p_i^2=1\,. \ee
In other words, the quadratic terms in the action (\ref{action}) do
not modify the Kasner conditions that arise in Einstein gravity.
This disallows isotropic expansion and, in particular, one exponent
must be negative. However, in conformal gravity the exponents need
only satisfy the condition
\be 2\sum_{i=1}^3 p_i +2\sum_{i=1}^3 p_i^2-\left( \sum_{i=1}^3
p_i\right)^2=3\,. \ee
This solution includes the Kasner metric for which both conditions
in (\ref{conditions}) are obeyed.

\subsection{pp-wave metrics}

A general class of pp-wave solutions for the general quadratic
action (\ref{action}) with arbitrary $\alpha$ and $\beta$ has the
metric
\be
ds^2=H dx^2+\fft{dr^2}{r^2}+r^2 (-2dt dx+dy^2)\,,
\ee
where
\be
H=f_1r^2+\fft{f_2}{r}+f_3r^{2z_+}+f_4r^{2z_-}+g_1(1+y^2r^2)\,,
\ee
the $f_i$ and $g_i$ are functions of $x$ only and $z=z_\pm$ satisfy
the equation
\be
1-24\beta+\alpha (4z_{\pm}^2-2z_{\pm}-8)=0\,.
\ee

For conformal gravity and critical gravity, the $H$ function can have
additional terms. Namely, for conformal gravity $H$ has the form
\be
H=f_1 r^2+\fft{f_2}{r}+f_3r+f_4+g_1y^2r^2+g_2y^3r^2\,,
\ee
while for critical gravity it is given by
\be
H=f_1r^2+\fft{f_2}{r}+f_3r^2\log r+f_4\fft{\log r}{r}+g_1(1+y^2r^2)\,.
\ee
For $g_i=0$ these metrics all reduce to ones presented in
\cite{aflog,ggst}, for which $H$ is a function only of $x$ and $r$.
Metrics for which the $H$ function involves sinusoidal dependence on
the $y$ coordinate are also discussed in \cite{ggst}. For
$f_1=f_3=f_4=g_i=0$ all of these metrics reduce to the Kaigorodov
\cite{kaig} metric.

\section{Energy of Logarithmic Black Hole in Critical Gravity}

In this appendix, we derive the mass of the logarithmic black hole
using the Abbott-Deser-Tekin (ADT) and the Ashtekar-Magnon-Das (AMD)
procedures.

The main idea of the ADT method is to write the asymptotic AdS
black hole metric in the form $g_{\mu\nu}=\bar g_{\mu\nu} +
h_{\mu\nu}$, where $\bar g_{\mu\nu}$ is the metric on AdS, and
then interpret the linearised variation of the field equation, given
in our case by (\ref{eom}), as an effective gravitational
energy-momentum tensor $T_{\mu\nu}$ for the black hole field. One
then writes the conserved current $J^\mu=T^{\mu\nu}\,\xi_\mu$, where
$\xi^\mu$ is a Killing vector that is timelike at infinity, as the
divergence of a 2-form ${\cal F}_{\mu\nu}$; {\it i.e.} $J^\mu=
\nabla_\nu {\cal F}^{\mu\nu}$. From this, one obtains the ADT mass
for the Lagrangian corresponding to (\ref{action}):
\be 8\pi GE =(1+8\Lambda\beta+2\Lambda\alpha)\, \int_{S_\infty}\,
dS_{i}\, {\cal F}_{(0)}^{0i} +(2\beta+\alpha) \int_{S_\infty}\,
dS_{i}\, {\cal F}_{(1)}^{0i}+\alpha \int_{S_\infty}\, dS_{\mu\nu}\,
{\cal F}_{(2)}^{0i}\,, \label{ADT} \ee
where $dS_{i}$ is the area of the  sphere at infinity. The
definition of ${\cal F}^{\mu\nu}$ associated with the various terms
in the equations of motion have been calculated in \cite{destek}.
One may verify that upon defining
\bea {\cal F}_{(0)}^{\mu\nu} &=& \xi_\alpha \nabla^{[\mu}
h^{\nu]\alpha}
 + \xi^{[\mu}\nabla^{\nu]}\, h + h^{\alpha[\mu}\nabla^{\nu]}\xi_\alpha
 -\xi^{[\mu}\, \nabla_\alpha h^{\nu]\alpha} + \ft12 h \nabla^\mu\xi^\nu\,,
 \nn\\
{\cal F}_{(1)}^{\mu\nu} &=& 2\xi^{[\mu}\, \nabla^{\nu]}\, R^L +
                                 R^L\, \nabla^\mu\xi^\nu\,,\nn\\
{\cal F}_{(2)}^{\mu\nu} &=& -2\xi_\alpha\, \nabla^{[\mu}\,
\cG_L^{\nu]\alpha}
 -2 \cG_L^{\alpha[\mu}\ \nabla^{\nu]}\, \xi_\alpha\,,
\eea
it follows that
\bea
\nabla_\nu\, {\cal F}_{(0)}^{\mu\nu} &=& \cG_L^{\mu\nu}\, \xi_\nu\,,\nn\\
\nabla_\nu\, {\cal F}_{(1)}^{\mu\nu} &=&\Big[(-\nabla_\mu\nabla_\nu
+
   g^{\mu\nu}\square + \Lambda\, g^{\mu\nu}) R^L\Big] \xi_\nu\,,\nn\\
\nabla_\nu\, {\cal F}_{(2)}^{\mu\nu} &=&
  \Big[(\square - \fft{2\Lambda}{3})\cG_L^{\mu\nu} -
   \fft{2\Lambda}{3}\, R^L\, g^{\mu\nu}\Big]\, \xi_\nu\,.
\eea
At the critical point $\Lambda\alpha=-3\Lambda\beta=\ft{3}{2}$, the
first term in (\ref{ADT}) vanishes, and the contributions to the
mass of the logarithmic black hole from the two remaining terms is
\begin{equation}\label{}
    E_{\rm{log}}^{\rm ADT}=\frac{3m}{8\pi G}\,.
\end{equation}

Since the logarithmic black hole is asymptotically AdS, one can also
try to apply the AMD method to this case. The derivation
of AMD conserved quantities relies on a detailed analysis of the
fall-off rate of the curvature near the boundary, which is weighted
by a smooth function $\Omega$ (the conformal boundary is defined at
$\Omega=0$). For details on the requirement for $\Omega$, the reader
is referred to \cite{Ashtekar:1984zz, Ashtekar:1999jx}. For
$n$-dimensional asymptotic AdS spacetime, for generic cases in which
the leading fall off of the Weyl tensor goes as $\Omega^{n-5}$, the
AMD formula for conserved quantities in quadratic curvature theories
were explored in \cite{Okuyama:2005fg, Pang:2011cs}. However, in the
case of AdS logarithmic black holes, the leading fall off of the
Weyl tensor near the boundary is modified to be
\begin{equation}\label{}
C_{abcd}\rightarrow \Omega^{n-5}K_{abcd}+\Omega^{n-5}\log(\Omega)
L_{abcd}\,.
\end{equation}
Here $a$ and $b$ are indices related to a new coordinate system
which adopts $\Omega$ as the radial coordinate. It is found that, at
critical points where a logarithmic term can appear, the fall-off
behavior of the energy-momentum tensor is still at the order of
$\Omega^{n-3}$. Thus, the flux across the boundary is finite. This
implies that the AMD conserved quantities for the logarithmic black hole
may be well defined. In \cite{Pang:2011cs}, the AMD conserved
quantities corresponding to the logarithmic black hole are found to be
given by
\begin{equation}
Q_{\xi}[C] = \frac{\alpha R_0}{8\pi G_{(n)}n(n-3)} \int_C dx^{n-2}
\sqrt{\hat{\sigma}} \hat{\cal L}_{ab} \xi^{a} \hat{N}^{b}\,,
\end{equation}
with
\begin{equation}\label{}
     R_0 =-n(n-1)\,,
\end{equation}
where the AdS radius has been set to 1 and $\hat{\cal L}_{ab}\equiv
\ell^2L_{eafb}{\hat n}^{e}{\hat n}^{f}$. Specifically for the
four-dimensional AdS-logarithmic black hole solutions with the
asymptotic expansion given by (\ref{log-bh}), one finds that
\begin{equation}\label{}
    E_{\rm{log}}^{\rm AMD}=\frac{3m}{8\pi G}\,.
\end{equation}

\section{Energy of AdS Black Holes in Conformal Gravity}

In this appendix, we present details of the proposal for calculating
the mass of AdS black holes in conformal gravity that we discussed
in section 5.1.
The Lagrangian for conformal gravity is given by
\begin{equation}
e^{-1}{\cal L}=\frac{\alpha}{2} C^{\mu\nu\rho\lambda}C_{\mu\nu\rho\lambda}\,.
\end{equation}
The solutions of conformal gravity discussed in section 5 can be written as
 \begin{equation}\label{sc}
ds^2=-f dt^2 + \fft{dr^2}{f} + r^2 d\Omega_{2,k}^2\,,\qquad f= r^2 +
b r + c + \fft{d}{r},\qquad 3bd-c^2+k^2=0\,,
\end{equation}

To apply the ADT method to this solution, a background subtraction
is necessary. It turns out that if we simply choose the static AdS
metric as the background, the energy calculated is divergent. On the
other hand, the background-independent AMD method will also give a
infinite result since the leading fall-off of the Weyl tensor of the
solution (\ref{sc}) is slower than that of the usual AdS black hole.

Motivated by finding a proper definition of energy for black hole
solutions (\ref{sc}) in conformal gravity, we adapt the standard
Noether method to the Lagrangian of conformal gravity. In the
following, we briefly review the Noether procedure for deriving a
conserved current associated with symmetry generated by the Killing
vector $\xi$.

The first variation of the Lagrangian 4-form generated by the
vector $\xi$ can always be expressed as
\begin{equation}
  {\cal L}_{\xi}L=E{\cal L}_{\xi}\phi+d\Theta(\phi,{\cal L}_{\xi}\phi),
\end{equation}
where $\phi$ represents a collection of tensorial fields, $E$ denotes
their equations of motion, and ${\cal L}_\xi$ denotes the Lie derivative.
Using the identity
\begin{equation}
    {\cal L}_{\xi}=di_{\xi}+i_{\xi}d,
\end{equation}
for the Lie derivative of a differential form,
we find a conserved current defined by
\begin{equation}
    J=\Theta-i_{\xi}L.
\end{equation}
On shell, we have
\begin{equation}
    dJ=0\Rightarrow J=dQ,
\end{equation}
where $Q$ is the conserved charge density associated with the symmetry
generated by $\xi$.
Applying this procedure to conformal gravity, we find that
\begin{equation}
   Q=\frac{1}{4}\epsilon_{\mu\nu\rho\sigma}Q^{\rho\sigma}dx^{\mu}\wedge
dx^{\nu},
\end{equation}
with
\begin{equation}\label{Qcf}
Q^{\rho\sigma} = -\frac{\alpha}{8\pi
G}(C^{\rho\sigma\mu\nu}\nabla_{\mu}\xi_{\nu}-2\xi_{\nu}
\nabla_{\mu}C^{\rho\sigma\mu\nu}).
\end{equation}
It is well known that the conserved charge $Q$ derived from
the Einstein-Hilbert action only accounts for one half of the true
ADM mass. (The other half can be understood as coming from a total
derivative term added to the Einstein-Hilbert action \cite{katz}.)
The validity of the proposal to take (\ref{Qcf}) as the definition
of energy for black holes in conformal gravity can be tested
by applying it to the known
examples of the Schwarzschild-AdS and Kerr-AdS black holes. We find that
the results using (\ref{Qcf}) coincide with those obtained from the
AMD method and in particular,
by setting $\alpha=\ft{1}{2}$, we recover the result presented in
\cite{Caldarelli:1999xj} and \cite{Gibbons:2004ai}, thus confirming
the tree level equivalence between Einstein gravity and Weyl gravity that
was proposed in \cite{mald}.

Finally, we calculate the conserved charge for the metric (\ref{sc})
associated with the timelike Killing vector $\partial/\partial t$.
It is given by
\begin{equation}
\int Q=-\frac{\alpha\omega_2}{16\pi G}[4d-\ft{2}{3}b(c-k)]\,.
\end{equation}
As we discuss in section 5, we can use this conserved quantity to
provide a definition of energy, which turns out to be consistent
with the first law of thermodynamics for the general AdS black holes
in conformal gravity.

\end{document}